\documentclass[aps,prd,reprint,superscriptaddress,nofootinbib,nobibnotes,notitlepage,floatfix]{revtex4-2}

\pdfoutput=1

\usepackage[T1]{fontenc}
\usepackage{graphicx}
\usepackage{amsmath}
\usepackage{amssymb}
\usepackage{bm}
\usepackage{xcolor}
\usepackage[colorlinks=true,linkcolor=blue!60!black,citecolor=blue!60!black,urlcolor=blue!60!black]{hyperref}
\usepackage[nameinlink]{cleveref}
\crefname{equation}{Eq.}{Eqs.}
\Crefname{equation}{Equation}{Equations}
\crefname{figure}{Fig.}{Figs.}
\Crefname{figure}{Figure}{Figures}
\crefname{table}{Table}{Tables}
\Crefname{table}{Table}{Tables}
\crefname{section}{Sec.}{Secs.}
\Crefname{section}{Section}{Sections}
\crefname{appendix}{Appendix}{Appendices}
\Crefname{appendix}{Appendix}{Appendices}
\crefname{subappendix}{Appendix}{Appendices}
\Crefname{subappendix}{Appendix}{Appendices}

\newcommand{\dd}{\mathrm{d}}
\newcommand{\Tcmb}{T_{\rm CMB}}

\begin{document}
\title{Detection of the Galaxy--Galaxy--Temperature Skewed Bispectrum}
\author{I.~J.~Keja}
\affiliation{Van Swinderen Institute for Particle Physics and Gravity, University of Groningen, 9747 AG Groningen, The Netherlands}
\author{D.~Karagiannis}
\email[Contact author: ]{dakaragian@gmail.com}
\affiliation{Van Swinderen Institute for Particle Physics and Gravity, University of Groningen, 9747 AG Groningen, The Netherlands}
\affiliation{Department of Physics \& Astronomy, University of the Western Cape, Cape Town 7535, South Africa}
\author{W.~R.~Coulton}
\affiliation{Department of Physics, University of Oxford, Denys Wilkinson Building, Keble Road, Oxford OX1 3RH, United Kingdom}
\author{P.~D.~Meerburg}
\affiliation{Van Swinderen Institute for Particle Physics and Gravity, University of Groningen, 9747 AG Groningen, The Netherlands}

\begin{abstract}
We present the first measurement of the galaxy--galaxy--temperature skewed bispectrum, obtained by cross-correlating the squared, filtered unWISE galaxy overdensity with \textit{Planck} maps at 100, 143, 217, and 353~GHz. The signal is detected with high significance in all four channels. Its frequency dependence reveals contributions from hot gas through the thermal Sunyaev--Zel'dovich (tSZ) effect and from dust through the cosmic infrared background (CIB). With the galaxy occupation fixed, our halo-model fits favor higher integrated pressure and peak star-formation efficiency for the galaxy--temperature cross-spectra than for the skewed bispectrum. Comparisons with component-separated tSZ and CIB maps from the same \textit{Planck} observations indicate that this mismatch is not confined to the frequency-map fits. This difference motivates testing more flexible descriptions of gas pressure, CIB emission, and galaxy occupation. In particular, it could point to a mass dependence of the gas-pressure profile that our model does not fully capture. By giving greater weight to massive halos that host many galaxy pairs, the skewed bispectrum provides a complementary test of these descriptions, using information beyond two-point statistics at a comparable computational cost.
\end{abstract}
\maketitle

\section{Introduction}

Two-point statistics of the cosmic microwave background (CMB) and large-scale structure have provided precise constraints on cosmology and on the properties of galaxies and gas. On nonlinear scales, however, they capture only part of the available information. Although the primordial fluctuations are nearly Gaussian \cite{2020_inflation_constraints}, gravitational evolution couples Fourier modes and generates a non-zero bispectrum. Higher-order statistics
are now being used in analyses of galaxy clustering \cite{Philcox_2022,Ivanov:2023qzb,DAmico:2022gki,Cagliari:2025rqe,
Chudaykin2025DESI1,Chudaykin2025DESI2,Chudaykin2025DESI3}
and weak lensing \cite{Gatti_2022_WL,Secco_2022,DES:2025llp,Sugiyama:2025dfc,Linke:2026vam}, and mixed bispectra of CMB lensing with the
Lyman-$\alpha$ forest \cite{Doux_2016,Karacayli:2024udo} and with galaxy pairs \cite{Farren_2023} have been detected.

Microwave observations probe the gas and dust associated with galaxies through the thermal Sunyaev--Zel'dovich (tSZ) effect and the cosmic infrared background (CIB). The tSZ effect traces the pressure of ionized gas and changes sign near 217~GHz \cite{Sunyaev_1972,Carlstrom_2002}, whereas the CIB is produced by dust in star-forming galaxies and rises steeply toward higher frequencies \cite{Maniyar_2021}. Galaxy cross-correlations have used these signals to study the thermal history of the gas and dusty star formation \cite{Koukoufilippas_2019,Chiang_2020,Gatti_2022_tSZ,Yan_2024}. Beyond such two-point measurements, the tSZ and CIB bispectra have been measured with \textit{Planck} \cite{2014_CIB_colorcorr}, the South Pole Telescope \cite{Crawford:2013uka}, and the Atacama Cosmology Telescope \cite{Coulton:2017crj}. The tSZ power spectrum and bispectrum are both sensitive to the amplitude of matter fluctuations \cite{Komatsu_2002,Hill_2013,Bhattacharya:2012nw}, and for South Pole Telescope data, adding the bispectrum halved the uncertainty on the tSZ power-spectrum amplitude \cite{Crawford:2013uka}.

Correlating microwave emission with galaxy pairs provides a further test of how galaxies, gas, and dust share their host halos. Stacking \textit{Planck} Compton-$y$ maps on pairs of luminous red galaxies has revealed tSZ emission from the filaments between them \cite{Tanimura:2017ixt,deGraaff:2017byg}. In the one-halo contribution, the galaxy--temperature cross-spectrum weights the emission by the mean number of galaxies in each halo. The galaxy--galaxy--temperature bispectrum instead weights it by the mean number of distinct galaxy pairs \cite{COORAY_2002}. This gives more weight to halos rich in galaxies, especially massive clusters. Because the two statistics weight the same halo population differently, a model that matches the galaxy--temperature cross-spectrum need not also match the bispectrum. Comparing them tests our assumptions about how galaxies populate halos and about the gas and dust within them.

This comparison does not require estimating every allowed angular-bispectrum triangle. We instead filter and square the projected galaxy overdensity and cross-correlate it with the microwave temperature field. The resulting skewed bispectrum sums over the two galaxy multipoles, retaining the dependence on the temperature multipole $L$. This filter-square, or skew, spectrum \cite{Dore_2004,Schmittfull:2014tca,MoradinezhadDizgah:2019xun,Chakraborty_2022,Chen:2024bdg,Harscouet_2025,Harscouet_2026} retains scale dependence, at a computational cost comparable to that of a power spectrum. Ref.~\cite{Farren_2023} used it to detect the unWISE--unWISE--CMB-lensing bispectrum. Here we investigate whether the same approach, applied to microwave temperature maps, can probe the hot gas and dust associated with galaxies.

Previous three-point analyses involving galaxies and microwave temperature have largely targeted the kinetic Sunyaev--Zel'dovich (kSZ) effect. One approach correlates galaxies with squared, filtered temperature maps, measuring a temperature--temperature--galaxy bispectrum \cite{Hill_2016,Ferraro:2016ymw,Kusiak:2021hai}. Another combines small-scale galaxy and temperature fluctuations with galaxy redshift information to reconstruct the large-scale line-of-sight velocity field. Correlating this reconstruction with galaxy density probes a galaxy--galaxy--temperature bispectrum that couples large- and small-scale modes \cite{Smith:2018bpn}. Recent analyses have correlated the kSZ-reconstructed velocity field with galaxy density \cite{Hotinli_2025_velocity,Lai_2025_velocity,Chaussidon:2026vmn} and with velocities predicted from that density field \cite{McCarthy:2024nik,McCarthy_2025_velocity}. The tSZ effect and the CIB are contaminants in these kSZ measurements \cite{Bloch:2024kpn,EmbilVillagra:2026fnh}, and galaxy tracers such as unWISE have been proposed for removing them from CMB maps \cite{Kusiak_2023}. Our statistic instead uses the squared projected galaxy field to target their contributions directly.

In this work, we present the first measurement of the galaxy--galaxy--temperature skewed bispectrum, using the unWISE Blue galaxy sample and \textit{Planck} HFI maps at 100, 143, 217, and 353~GHz. We detect it in all four channels and test it with polarization null maps and alternative sky masks. We fit its frequency and scale dependence with a halo model, using a covariance that includes fluctuations in the halo population, and compare the inferred gas and dust parameters with those from the galaxy--temperature cross-spectra at fixed galaxy occupation. The two statistics favor different parameters, and we use component-separated maps to investigate this difference.

The Letter is structured as follows. \Cref{sec:data,sec:estimator} describe the data and the skewed-bispectrum estimator, and \cref{sec:detect} presents the detection and robustness tests. \Cref{sec:halomodel,sec:covariance} introduce the halo model and the covariance used for parameter inference. \Cref{sec:params} gives the constraints from the skewed bispectrum, and \cref{sec:crossfit} compares them with those from the cross-spectra and the component-separated maps. We conclude in \cref{sec:conclusions}. \Crefrange{sec:maps}{sec:components} provide further details of the data, methods, and tests, together with supporting results.

\section{Data}
\label{sec:data}

We use the unWISE Blue galaxy sample \cite{Schlafly_2019,Krolewski_2020}, selected by W1--W2 color from WISE and NEOWISE imaging \cite{Wright_2010}. It has mean redshift $\bar z\simeq0.6$ and width $\delta_z\simeq0.3$ \cite{Krolewski_2020,Kusiak_2022}, and contains approximately $8\times10^7$ galaxies within the analysis mask. We correct the number counts with the stellar-density and depth weight maps of Ref.~\cite{Krolewski_2022}, normalize them by the effective unmasked area of each pixel \cite{Krolewski_2020}, and form the overdensity $g=n/\bar n_g-1$ (\cref{eq:overdensity}), where $n$ is the corrected galaxy surface density per steradian and
$\bar n_g$ is its mean over the retained galaxy footprint.

The microwave fields $T_\nu$ are dimensionless temperature fluctuations obtained by dividing the \textit{Planck} PR3 full-mission HFI maps at 100, 143, 217, and 353~GHz \cite{2020_Planck_HFImaps} by $\Tcmb$. We combine the GAL060 Galactic mask with the point-source mask of each frequency \cite{2016_Planck_GalMasksPS} (\cref{sec:maps}). All maps are analyzed at $N_{\rm side}=2048$ after subtracting their masked means. 

\section{Estimator}
\label{sec:estimator}

On the full sky, the skewed-bispectrum estimator is
\begin{equation}
 \hat C_L^{g^2T_\nu}\equiv\frac{1}{2L+1}\sum_M a_{LM}^{g^2}a_{LM}^{T_\nu*},
 \label{eq:skew}
\end{equation}
where $g^2$ is the square of the galaxy overdensity after filtering with $w_\ell$ (\cref{sec:ranges}). We filter the galaxy field before squaring so that excluded galaxy modes do not contribute to the statistic. Its expectation value $C_L^{g^2T_\nu}$ is related to the angle-averaged angular bispectrum $B^{ggT_\nu}_{\ell_1\ell_2L}$ (modeled in \cref{sec:compression}) by
\begin{equation}
 C_L^{g^2T_\nu}=\sum_{\ell_1\ell_2}w_{\ell_1}w_{\ell_2}
 \sqrt{\tfrac{(2\ell_1+1)(2\ell_2+1)}{4\pi(2L+1)}}
 \begin{pmatrix}\ell_1&\ell_2&L\\0&0&0\end{pmatrix}
 B^{ggT_\nu}_{\ell_1\ell_2L}.
 \label{eq:compression}
\end{equation}
At each temperature multipole $L$, this expression sums over the two galaxy multipoles $\ell_1$ and $\ell_2$ selected by the filter \cite{Farren_2023,Harscouet_2025}. The compression retains scale dependence but does not
preserve the full dependence on triangle shapes.

We measure bandpowers of width $\Delta L=100$, each labelled by its central multipole $L$, with \textsc{NaMaster} \cite{Hivon_2002,Alonso_2019}, using a separate mode-coupling matrix for each mask pair and forward-modeling the beam, pixel, and bandpower windows. For detection S/N and parameter inference, we use the bandpowers with $600\leq L<2000$ at 100 and 143~GHz and $300\leq L<2500$ at 217 and 353~GHz. The galaxy filter is the same in all channels. For comparison, we measure the galaxy--temperature cross-spectra $C_L^{gT_\nu}$ using the same procedure and multipole ranges. This gives 72 bandpowers across the four frequency channels for each statistic (\cref{sec:ranges}).

\section{Detection and robustness}
\label{sec:detect}

We test the null hypothesis that the microwave temperature field is statistically independent of the galaxy field, so that its cross-correlations with both $g$ and $g^2$ vanish on average. Following Ref.~\cite{Farren_2023}, we use the covariance under the null hypothesis for the detection statistics. We estimate this covariance, $\widehat{\sf C}_0$, by cross-correlating the observed $g$ and $g^2$ with 300 \textit{Planck} Full Focal Plane 10 (FFP10) CMB-plus-noise
realizations per channel \cite{2020_Planck_HFImaps}, to which we add simulated foregrounds that are uncorrelated with the galaxies but have the measured auto- and cross-frequency power. The resulting errors agree with the Gaussian (Knox) expression \cite{Knox:1995dq} evaluated with the measured spectra (\cref{eq:knox,fig:errors}).

\Cref{tab:summary} reports the S/N and detection significance for both statistics. The S/N measures the size of the signal relative to the noise,
whereas the detection significance describes how unlikely it
would be to obtain a signal at least this strong if the galaxy and temperature fields were uncorrelated. We calculate the S/N using the Hartlap-corrected inverse null
covariance \cite{Hartlap_2006} and express the probability of obtaining such a signal as an equivalent one-sided Gaussian significance, in units of $\sigma$. This probability depends on the number of bandpowers combined
and on the uncertainty in the covariance estimated from 300
simulations. The Hartlap correction removes the average bias
in the inverse covariance, but this uncertainty remains.
Accounting for both effects gives detection significances
lower than the S/N values for these measurements. The difference grows at high S/N, where the probabilities of rare fluctuations are particularly sensitive to covariance uncertainty (\cref{eq:hotelling}).

The skewed bispectrum reaches S/N of 20 to 77 per channel and 110 jointly, with significances above $13\sigma$ in every channel. Combining the channels raises the S/N but also increases the number of bandpowers, from 22 at 353~GHz alone to 72, and with it the uncertainty of their covariance estimated from the same 300 simulations. For these measurements, the greater covariance uncertainty outweighs the gain in S/N, giving a lower joint significance than at 353~GHz. For $g^2T$, the S/N rises from 77 to 110, but the significance decreases from $28.0\sigma$ to $27.3\sigma$. Likewise, for $gT$ the significance decreases from $35.5\sigma$ to $31.9\sigma$.

Over the same multipole ranges, the $gT$ cross-spectra reach S/N of 12 to 184 per channel and 200 jointly. In the tSZ-dominated channels at 100 and 143~GHz, the skewed bispectrum has the higher S/N, 21 and 24, compared with 16 and 12 for the cross-spectra. Within each channel, the temperature spectrum and number of modes are the same, so the errors differ mainly through the galaxy auto-spectra $C_L^{gg}$ and $C_L^{g^2g^2}$. At 100 and 143~GHz, the $g^2T$ signal is larger relative to its null errors, giving it a higher S/N and significance than $gT$. These S/N values use the null covariance. For parameter inference, we adopt a more complete covariance that depends on the fitted emission parameters and accounts for fluctuations in the halo population. This additional source of uncertainty particularly affects \(g^2T\) and does not enter the detection S/N (\cref{sec:clusterterm}).

\Cref{fig:g2t-fit} shows the frequency dependence. The skewed bispectrum is negative at 100 and 143~GHz, where the tSZ contribution dominates, and positive at 217 and 353~GHz, where CIB emission dominates. At 217~GHz, the tSZ contribution nearly vanishes. This sign reversal and the different scale dependence of the two components allow both to be fitted directly to the frequency maps, without first constructing component maps.

Previous unWISE--\textit{Planck} temperature cross-correlations detected the large-scale integrated Sachs--Wolfe effect at $3.2\sigma$ across three galaxy samples \cite{Krolewski_2022}. On smaller scales, analyses of lower-redshift samples have separated the tSZ, CIB, and radio contributions \cite{Zhao_2026}. Published galaxy correlations with \textit{Planck} component maps reach S/N of 22 for WISE galaxies with the Compton-$y$ map \cite{Ibitoye:2022oot} and 194 for three unWISE samples with the CIB maps at 353--857~GHz \cite{Yan_2024}, comparable to those of our cross-spectra. Similarly, the S/N of 21 and 24 that the skewed bispectrum reaches at 100 and 143~GHz is comparable to the S/N of 26 for the unWISE Blue galaxy--galaxy--CMB-lensing bispectrum \cite{Farren_2023}. These comparisons are only indicative, because the galaxy samples, sky coverage, multipole ranges, galaxy filters, noise, and covariance estimates differ.

\begin{table}[t]
\caption{Detection and robustness tests over the analysis ranges. Top: S/N and detection significance (\cref{eq:snr,eq:hotelling}). Middle: PTEs of the polarization null tests. Small PTEs indicate disagreement with zero. All tests shown are consistent with zero. Bottom: percentage change of the overall $g^2T$ amplitude relative to the fiducial measurement. GAL masks are named after the percentage of sky they retain. Whether the changes across the multipole bins are consistent with CMB and noise fluctuations is tested in \cref{tab:null_summary}.}
\label{tab:summary}
\centering
\begin{ruledtabular}
\begin{tabular}{lccccc}
$\nu$ [GHz] & 100 & 143 & 217 & 353 & Joint \\
\hline
\multicolumn{6}{l}{\textit{Detection}} \\
$gT$ S/N & 16 & 12 & 61 & 184 & 200 \\
$gT$ significance [$\sigma$] & 11.9 & 8.5 & 25.8 & 35.5 & 31.9 \\
$g^2T$ S/N & 21 & 24 & 20 & 77 & 110 \\
$g^2T$ significance [$\sigma$] & 14.6 & 16.3 & 13.3 & 28.0 & 27.3 \\
\hline
\multicolumn{6}{l}{\textit{Polarization null tests}} \\
$g^2E$ PTE & 0.58 & 0.25 & 0.79 & 0.84 & 0.85 \\
$g^2B$ PTE & 0.18 & 0.79 & 0.31 & 0.41 & 0.28 \\
\hline
\multicolumn{6}{l}{\textit{Change of $g^2T$ amplitude with sky mask [\%]}} \\
GAL020 & $+4.6$ & $+2.9$ & $+11.0$ & $+4.0$ & -- \\
GAL040 & $+1.6$ & $+1.2$ & $+3.4$ & $+3.2$ & -- \\
GAL070 & $-1.8$ & $-0.5$ & $-1.5$ & $-0.1$ & -- \\
Common mask & $-6.9$ & $-5.3$ & $-1.1$ & $-1.2$ & -- \\
\end{tabular}
\end{ruledtabular}
\end{table}

We perform a null test by replacing temperature with the \textit{Planck} $E$- and $B$-mode maps while keeping the galaxy fields unchanged. The dominant tSZ and CIB signals are effectively unpolarized, so these correlations should vanish. We quantify agreement with zero using the probability to exceed (PTE), the probability that random fluctuations alone would give a departure from zero at least as large as the one observed. All four spectra ($gE$, $gB$, $g^2E$, $g^2B$) are consistent with zero, with PTEs of 0.18--0.85 for $g^2E$ and $g^2B$, providing no evidence for polarized contamination correlated with the galaxies (\cref{tab:summary,sec:tests}).

We also repeat the measurement with alternative sky masks to test how the signal depends on the retained sky. At 100 and 143~GHz, which constrain the gas pressure, the overall amplitude changes by at most 6.9\% and 5.3\%, respectively, while at 217 and 353~GHz the high-latitude GAL020 and GAL040 masks increase it by 3--11\%. These overall changes can be smaller than those in individual multipole bins (\cref{fig:masks,tab:null_summary}). For example, adding Galactic sky with either GAL070 or the \textit{Planck} common intensity mask changes the 353~GHz amplitude by at most 1.2\%, yet increases the largest-scale bandpower ($300\leq L<400$) by 41\%. Since GAL070 retains the fiducial point-source masks, this excess also appears when only the Galactic mask is changed, suggesting additional Galactic dust in the added sky (\cref{sec:tests}). The polarization nulls do not exclude this interpretation, since weakly polarized dust can contribute to the temperature correlations without producing a detectable polarization signal.

\section{Halo model}
\label{sec:halomodel}

Our halo model \cite{COORAY_2002} combines a halo occupation distribution (HOD) for the galaxies with models for the CIB emission \cite{Maniyar_2021,Yan_2024} and the electron pressure \cite{Arnaud_2010,2013Planck_tSZparams_bestfit}. We project its one-, two-, and three-halo bispectra in the Limber approximation and convert them into the skewed bispectrum with \cref{eq:compression}, using the same filter $w_\ell$ as in the measurement (\cref{eq:compression_h}). The one-halo term weights each halo by its number of distinct galaxy pairs, so its tSZ part is dominated by massive clusters (\cref{fig:massweights}). The electrons in these clusters are hot enough for relativistic corrections to the tSZ spectrum to matter \cite{Chluba_2012,Remazeilles_2019}. We compute these corrections with SZpack \cite{Chluba_2012} and include them without a free parameter. The occupation, mass conventions, and emission profiles are specified in \cref{sec:profiles}. For the model fits, we measure the shot-noise term from the catalogue and temperature maps and subtract it from the skewed bispectrum, in both the data and in every simulation (\cref{sec:shotnoise}).

We first fit the five-parameter HOD to the unWISE Blue galaxy auto-spectrum over $100\leq L<2000$ (\cref{sec:profiles}). We then hold this galaxy occupation fixed and fit five parameters to the skewed bispectrum. The two CIB parameters of the star-formation efficiency (\cref{eq:cib_efficiency}) are the halo mass $M_{\rm eff}$ at which star formation is most efficient, fitted as $\log_{10}(M_{\rm eff}/M_\odot)$, and the peak efficiency $\eta_{\rm max}$, which gives the fraction of accreted baryons converted into stars at that mass \cite{Maniyar_2021}. For the gas pressure, we fit the normalization $P_0$ and concentration $c_{500}$ of the generalized Navarro--Frenk--White (NFW) profile \cite{Arnaud_2010} in \cref{eq:gnfw}.

Part of the CIB signal comes from dust emitted by the selected unWISE galaxies themselves. We call this the self-dust contribution. In $gT$, a galaxy correlates with its own emission, giving a scale-independent term before instrumental smoothing. In $g^2T$, one member of a distinct galaxy pair supplies the emission, so the contribution depends on the clustering of the other galaxy. It survives shot-noise subtraction, which removes only pairs in which the same galaxy is counted twice. The fifth parameter, $A_{353}^{g^2T}$, sets its amplitude at 353~GHz. Before beam smoothing and binning, the models for the cross-spectra and the skewed bispectra are
\begin{align}
 C_L^{gT_\nu}&=C_L^{gT_\nu,\,\mathrm{tSZ}}+C_L^{gT_\nu,\,\mathrm{CIB}}+\delta_{\nu,353}\,A_{353}^{gT},
 \label{eq:gt_cross_shot_model}\\
 C_L^{g^2T_\nu}&=C_L^{g^2T_\nu,\,\mathrm{tSZ}}+C_L^{g^2T_\nu,\,\mathrm{CIB}}+\delta_{\nu,353}\,A_{353}^{g^2T}\,\mathcal S_L,
 \label{eq:cib_coincident}
\end{align}
where the tSZ and CIB terms are the halo-model predictions (\cref{sec:model}) and the last term is the self-dust contribution, which $\delta_{\nu,353}$ restricts to 353~GHz. We tested scaling this term from 353~GHz with the CIB dust spectrum. The predicted contributions fall below the uncertainties at 100 and 143~GHz, while the 217~GHz $gT$ data disfavor this scaling
(\cref{sec:robustness}). In $g^2T$, its scale dependence $\mathcal S_L$ is predicted by the halo model (\cref{eq:own_dust_shape}). After beam and pixel-window smoothing and bandpower binning, these spectra form the model vector $\bm m(\bm\theta)$ of \cref{eq:complete_gaussian}.

We fix the outer pressure slope to the \textit{Planck} cluster value $\beta_{\rm tSZ}=4.13$ \cite{2013Planck_tSZparams_bestfit}. Since the pressure profile was calibrated using X-ray hydrostatic masses, we evaluate it at the hydrostatic mass $M_{500c}/B$, fixing the ratio of true to hydrostatic mass to $B=1.25$ \cite{Bolliet_2018,2014_Planck_SZcounts}.

\section{Covariance}
\label{sec:covariance}

For parameter inference, we use a Gaussian likelihood with covariance
\begin{equation}
\mathsf C(\bm\theta)=h_{\rm H}^{-1}\mathsf C_{\rm mock}+\mathsf C_{\rm halo}(\bm\theta),
 \label{eq:cov_split}
\end{equation}
where $\mathsf C_{\rm mock}$ describes microwave fluctuations at fixed observed galaxy maps, while $\mathsf C_{\rm halo}$ approximates the additional scatter from fluctuations in halo abundance and galaxy occupation.\footnote{This prescription matches the covariance of the marginalized
likelihood under the assumptions in \cref{sec:inference} \cite{Heavens_2026}.
We use a Gaussian approximation to that likelihood.} The Hartlap factor $h_{\rm H}$ is defined in \cref{eq:snr}. The mock covariance estimates the disconnected part of the covariance. For a skewed-bispectrum bandpower, the disconnected variance is proportional to $C^{g^2g^2}C^{TT}+(C^{g^2T})^2$ per mode. The second term is below 0.4\% of the first and is neglected (\cref{eq:knox}). We compute $\mathsf C_{\rm mock}$ from the same 300 temperature realizations as for the null covariance, using the shot-noise-subtracted bandpowers for the skewed bispectrum (\cref{eq:paired_cov}). Because the realizations are cross-correlated with the observed galaxy maps using the data masks and estimator, $\mathsf C_{\rm mock}$ includes mask effects, correlations between frequency channels, and the measured galaxy clustering and shot noise. It also retains the non-Gaussian correlations entering $C^{g^2g^2}$, without requiring galaxy simulations.

Holding the galaxy maps fixed does not sample variations in the halo population and its correlated microwave emission. The term $\mathsf C_{\rm halo}$ models these as Poisson fluctuations in the halo counts and scatter in the galaxy occupation. We compute it with the same fixed HOD as the signal model and let it vary with the fitted emission amplitudes and pressure profile, including tSZ--tSZ, tSZ--CIB, and CIB--CIB terms (\cref{sec:clusterterm}). We neglect additional non-Gaussian covariance contributions beyond this approximation, including mixed galaxy--temperature trispectrum terms, products of bispectra, and connected six-point terms. These omissions include contributions involving several halos and super-sample covariance from density fluctuations on scales larger than the survey. 

\begin{figure*}[t]
\includegraphics[width=0.92\textwidth]{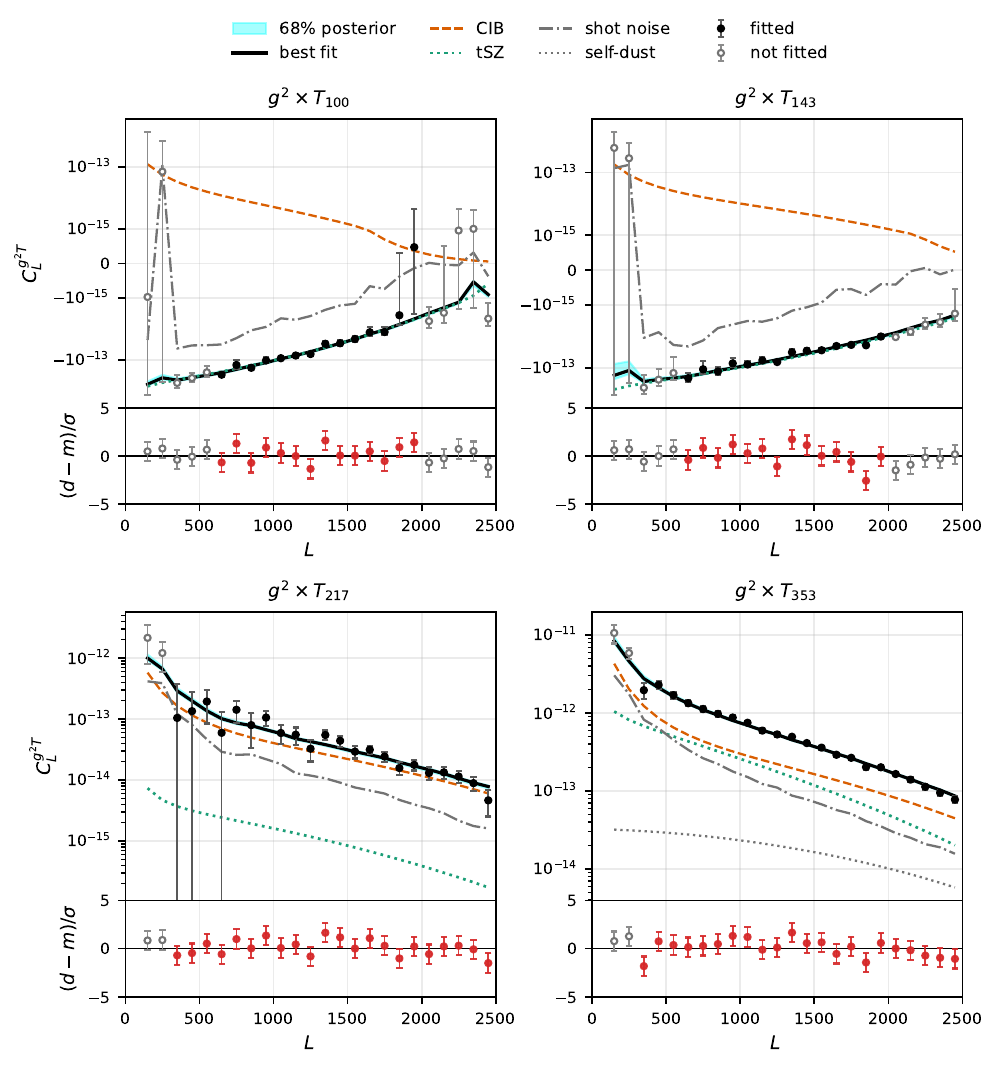}
\caption{Measured galaxy--galaxy--temperature skewed bispectra (points) and the best-fit halo model (black), split into CIB (orange dashed), tSZ (green dotted), the measured shot-noise term (gray dot-dashed), and the 353~GHz self-dust term at its fitted amplitude $A_{353}^{g^2T}$ (gray dotted). Both the points and the model include the instrumental beam and pixel window. Cyan bands contain 68\% of the posterior predictions. Filled points are fitted and open points are not. Lower panels show residuals in units of the diagonal uncertainties.}
\label{fig:g2t-fit}
\end{figure*}

\section{Parameter constraints}
\label{sec:params}

We sample the posterior with nested sampling \cite{Skilling_2006} using \textsc{dynesty} \cite{Speagle_2020}, and take the sample with the maximum likelihood as the best fit (\cref{sec:inference}). \Cref{fig:g2t-fit} compares the measured skewed bispectra with the best-fit halo model. A single set of five parameters captures the sign change between the tSZ- and CIB-dominated channels and the main scale dependence within each channel. At the best fit, the Hartlap-corrected reduced chi-squared is $\chi^2_{\rm red}=77.0/(72-5)=1.15$ (${\rm PTE}=0.19$). The best-fit parameters are
\begin{gather*}
(\log_{10}(M_{\rm eff}/M_\odot),\eta_{\rm max},P_0,c_{500},10^{12}A_{353}^{g^2T})\\
=(12.985,0.100,0.680,1.313,0.182).
\end{gather*}

\begin{figure*}[t]
\includegraphics[width=0.72\textwidth]{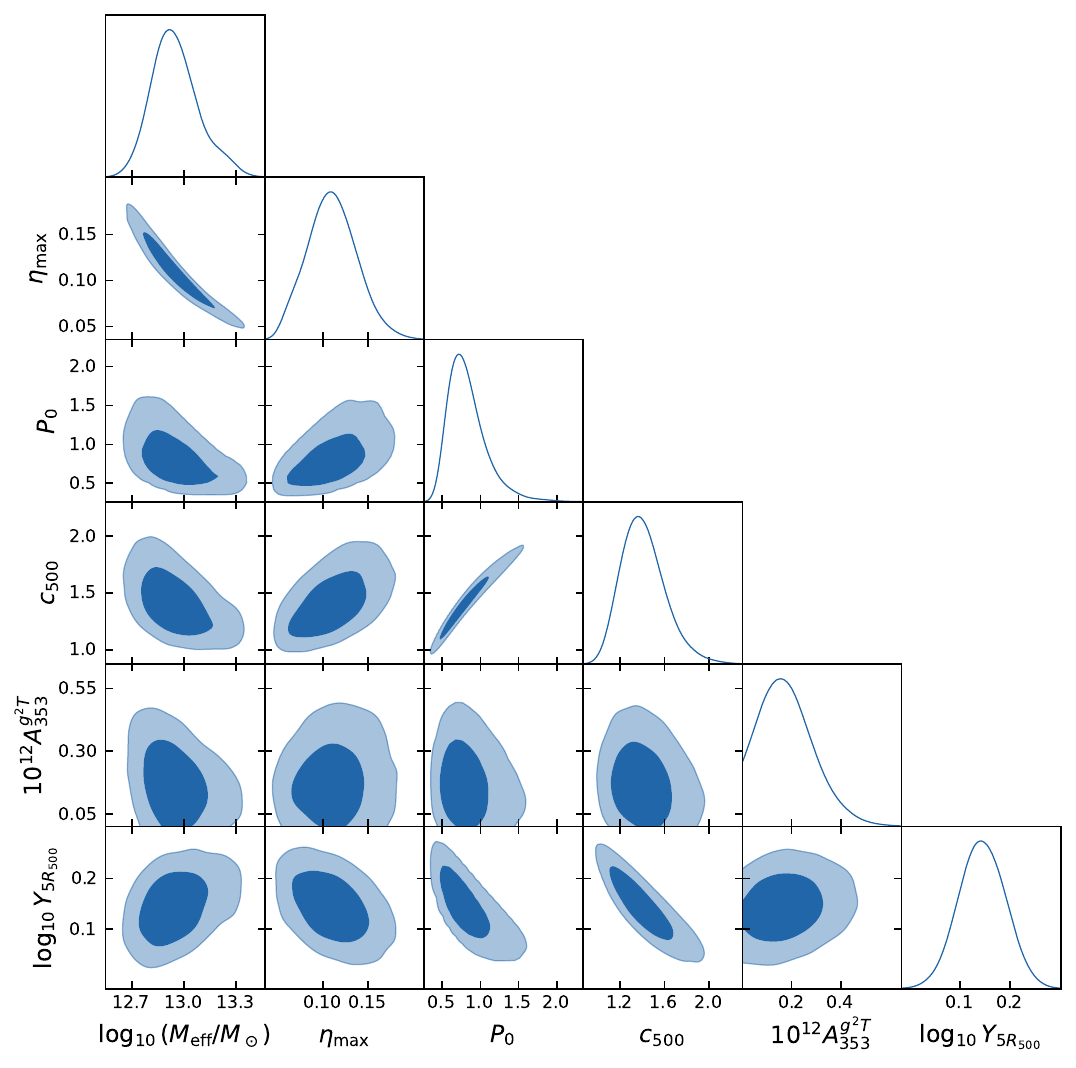}
\caption{Marginalized 68\% and 95\% posterior contours for the five free parameters of the $g^2T$ fit and the derived integrated pressure $\log_{10}Y_{5R_{500}}$. The outer pressure slope is fixed to $\beta_{\rm tSZ}=4.13$, and the galaxy occupation is fixed to the auto-spectrum fit. Priors and fixed parameters are listed in \cref{tab:g2params}.}
\label{fig:g2t-corner}
\end{figure*}

\Cref{fig:g2t-corner} shows the marginalized parameter constraints, while the priors and numerical results are listed in \cref{tab:g2params}. The contours show strong correlations between $M_{\rm eff}$ and $\eta_{\rm max}$ for the CIB, and between $P_0$ and $c_{500}$ for the gas pressure. In addition to the individual constraints on $P_0$ and $c_{500}$, we report a single integrated-pressure parameter, which combines them into a better-constrained quantity and allows comparison with integrated cluster pressure profiles. We define the dimensionless integrated pressure within $5R_{500}$ as
\begin{equation}
 Y_{5R_{500}}\equiv
 \frac{\int_0^{5R_{500}}4\pi r^2P_e(r)\,\dd r}
      {P_{500}R_{500}^3},
 \label{eq:pressure}
\end{equation}
where $P_e$ is the electron pressure and $P_{500}$ its
characteristic scale. The radius $R_{500}$ is computed
from the hydrostatic mass and encloses a mean density
500 times the critical density (\cref{sec:profiles}). Because the pressure profile scales with $P_{500}$ and $R_{500}$, $Y_{5R_{500}}$ depends only on the profile parameters, not on halo mass or redshift. We find
\begin{equation}
 \log_{10}Y_{5R_{500}}=0.144^{+0.048}_{-0.047},
 \label{eq:result}
\end{equation}
where we quote the posterior median and the $1\sigma$ uncertainty, given by the 16th and 84th percentiles. This corresponds to approximately 11\% precision
on $Y_{5R_{500}}$. 

Allowing the outer pressure slope to vary gives
$\log_{10}Y_{5R_{500}}=0.164^{+0.060}_{-0.066}$, consistent with the baseline result but with larger uncertainties
(\cref{fig:betafree}). Integrating the \textit{Planck} best-fitting pressure profile of nearby massive clusters \cite{2013Planck_tSZparams_bestfit} with \cref{eq:pressure} gives
$\log_{10}Y_{5R_{500}}=0.741$. At our adopted mass bias, this reference profile gives about four times the integrated pressure inferred from $g^2T$
(\cref{sec:constraints}).

\section{Comparison with the cross-spectra}
\label{sec:crossfit}

We fit the four galaxy--temperature cross-spectra with the same model, galaxy occupation, and covariance construction. In this fit, the fifth parameter, $A_{353}^{gT}$, is the amplitude of the 353~GHz self-dust term, which for $gT$ is a scale-independent term from the correlation of each galaxy with its own dust emission (\cref{eq:gt_cross_shot_model}). At the best fit, $\chi^2_{\rm red}=75.4/(72-5)=1.13$ (${\rm PTE}=0.22$), while the posterior gives
\[
 \log_{10}Y_{5R_{500}}=0.580^{+0.029}_{-0.032},
\]
corresponding to 7\% precision on $Y_{5R_{500}}$. This uncertainty is 1.6 times smaller than that of the skewed bispectrum, although the skewed bispectrum has the higher S/N in the tSZ-dominated channels (\cref{tab:summary}). The difference comes from $\mathsf C_{\rm halo}$, which enters the likelihood but not the null covariance used for the S/N. Because the skewed bispectrum weights halos by their galaxy pairs, its signal comes mainly from rare massive halos, and Poisson fluctuations in their number increase the uncertainty on the integrated pressure by about 53\%, compared with 3\% for the cross-spectra. Without $\mathsf C_{\rm halo}$, the two statistics would constrain $Y_{5R_{500}}$ equally well, to about 7\% (\cref{sec:clusterterm}).

The individual marginalized $1\sigma$ errors on $P_0$ and $c_{500}$ overlap between the fits, whereas those on the integrated pressure $Y_{5R_{500}}$, which combines these correlated parameters, do not. \Cref{fig:overlay} compares the parameter posteriors. The cross-spectra give 2.7 times more integrated pressure than the skewed bispectrum and a 1.9 times higher peak star-formation efficiency. The pressure difference between the two statistics persists when the 353~GHz channel is removed, when the 100 and 143~GHz ranges are extended to $L=400$, and when the outer pressure slope is free (\cref{sec:robustness}). At our adopted mass bias of $B=1.25$, the \textit{Planck} reference profile gives 1.4 times the integrated pressure inferred from $gT$. Increasing the mass bias lowers the pressure predicted for a halo of a given true mass. Keeping the reference profile shape fixed, we can therefore match our integrated pressures with mass biases of about 1.5 for $gT$ and 2.7 for $g^2T$ (\cref{eq:beq}). Previous \textit{Planck} tSZ analyses report central values between 1.3 and 1.7, using different data combinations and pressure models \cite{Koukoufilippas_2019,Koukoufilippas_2020_erratum,Ibitoye:2022oot,Bolliet_2018}. The equivalent value for $gT$ lies within this range, while that for $g^2T$ lies well above it. \Cref{sec:constraints} describes this amplitude conversion and compares our pressure and CIB results with previous analyses.

To test whether the mismatch depends on how the frequency-map fits separate gas and dust, we compare their predictions with maps in which the two components are separated directly, the \textit{Planck} NILC Compton-$y$ map \cite{2016Planck_ymap} and the 353~GHz CIB map of Ref.~\cite{Lenz_2019} (\cref{fig:component-validation}). Keeping the parameters of the frequency-map fits fixed, we compare the predicted cross-spectra and skewed bispectra with the component-map measurements. In both component maps, the $gT$ best fit overpredicts the skewed bispectrum relative to the cross-spectrum. Predictions from the $g^2T$ fit lie closer to the measured skewed bispectra. This suggests that the mismatch is not explained solely by how the frequency-map fits separate gas and dust. Because these comparisons use approximate errors and share the \textit{Planck} observations with the frequency-map fits, we treat them as descriptive checks (\cref{sec:components}).

The mismatch could arise if our occupation model places too many galaxy pairs in massive halos, making the predicted $g^2T$ signal too strong relative to $gT$. With fewer such pairs, the $gT$ fit would predict a lower skewed bispectrum, closer to the component-map measurements, and a refit of $g^2T$ would favor a higher pressure. The galaxy auto-spectrum and the galaxy--galaxy--CMB-lensing bispectrum also depend on galaxy pairs. For the same galaxies, Ref.~\cite{Farren_2023} found that the lensing bispectrum agrees reasonably well, particularly on large scales, with a halo-model prediction from an occupation fitted to two-point spectra \cite{Kusiak_2023}. However, the auto-spectrum counts all pairs equally and lensing weights them by halo mass, whereas the skewed bispectrum weights them by the tSZ signal of their halo, which rises more steeply with mass. Too many pairs in the most massive halos would therefore affect the skewed bispectrum much more than the other two statistics. Our microwave fits cannot correct this because they hold the occupation fixed at its auto-spectrum fit (\cref{tab:cggparams}). A joint fit of the galaxy auto-spectrum, cross-spectra, and skewed bispectrum, with the occupation and emission parameters free, would test this explanation and include the occupation uncertainty in the pressure and CIB constraints.

A different mass dependence of the gas-pressure profile could also contribute to the mismatch. If pressure in massive halos grows more slowly with mass than our model assumes (\cref{eq:p500}), the skewed-bispectrum fit could favor a lower $Y_{5R_{500}}$. Studies of tSZ correlations with weak lensing and galaxies have found lower pressure than predicted by some models in galaxy groups \cite{Pandey_2022_pressure,Pandey_2025_pressure,Siegel_2026_pressure}. These measurements probe lower halo masses than those dominating our modeled tSZ signal, so they provide context for testing pressure models across halo mass, rather than a direct explanation of the difference found here.

\section{Conclusions}
\label{sec:conclusions}

Using unWISE Blue galaxies and \textit{Planck} temperature maps, we have detected the galaxy--galaxy--temperature skewed bispectrum for the first time. The measurement reaches a joint S/N of 110 and contains contributions from both hot gas and dust. With the galaxy occupation fixed by the auto-spectrum, a halo model with five gas and dust parameters reproduces the measurement in all four channels and constrains the integrated pressure to 11\%. 

Applying the same model to the cross-spectra gives 2.7 times more integrated pressure and a higher peak star-formation efficiency. None of the changes to the frequency channels, multipole ranges, outer pressure slope, or added model terms that we tested removes this difference (\cref{sec:robustness}). Comparisons with component-separated maps, built differently from the same \textit{Planck} observations, support a difference in both gas and dust, so it does not arise only from how the frequency-map fits separate the two. A more complete treatment of the non-Gaussian covariance could increase
the parameter uncertainties, especially for the skewed bispectrum, and bring the two results into closer agreement.

The difference may indicate that our occupation model places too many galaxy pairs in massive halos, or that the gas-pressure profile varies with halo mass in a way that the adopted scaling does not capture. Either would affect the skewed bispectrum more than the cross-spectra, because it gives these halos more weight. Testing these possibilities will require fitting the occupation together with the emission. Simulations of the correlated galaxy, tSZ, and CIB fields would also help assess the covariance used in the fits. The low computational cost of the estimator makes it practical to repeat these measurements in other redshift ranges and with different filters, which would further test the difference.

Our results show that three-point correlations of galaxies with microwave temperature can be measured at high significance with existing data, and that they respond to the halo population differently from two-point statistics. Combining the two statistics, we could test the mass dependence of the gas pressure and thus models of baryonic feedback, which regulate galaxy formation and redistribute matter on the small scales used by cosmological surveys \cite{Chisari:2019tus,Troster:2021gsz}. Once the occupation and emission uncertainties are better controlled, the sensitivity of the skewed bispectrum to the abundance of massive halos could also provide information on the amplitude of matter fluctuations, as in analyses of the tSZ power spectrum and bispectrum \cite{Komatsu_2002,Bhattacharya:2012nw}. Upcoming galaxy surveys and CMB experiments with higher resolution and lower noise will sharpen these tests and extend their reach across cosmic time.

\begin{acknowledgments}
\begin{samepage}
DK is supported by the Dutch Research Council (NWO) (grant no. \href{https://www.nwo.nl/projecten/iolfn50333}{OCENW.M.22.307}). DK and PDM would like to thank the Center for Information Technology of the University of Groningen for their support and for providing access to the Hábrók high-performance computing cluster where the entire analysis was run. WRC is supported by the Science and Technology Facilities Council (STFC) through an Ernest Rutherford Fellowship (UKRI2424: ``Astrophysics and Cosmology from CMB Scattering'') and through the Royal Society grant ICA\textbackslash R2\textbackslash 252140. We thank Alex Krolewski, Gerrit Farren, Selim Hotinli and Abhi Maniyar for helpful discussions. We acknowledge the use of LLM agents to help with writing and editing, plotting, workflow management, and building further on analysis code written by DK. The authors orchestrated the whole process, checked all outputs, and take full responsibility for the scientific content. The analysis software is kept in a private Git repository and is available upon reasonable request. The assumptions and methodology implemented in the code are described in the appendices.
\end{samepage}
\end{acknowledgments}

\bibliographystyle{apsrev4-2}
\bibliography{references}

\appendix
\crefalias{section}{appendix}         
\crefalias{subsection}{subappendix}
\onecolumngrid

\section{Data, estimator, and simulations}
\label{sec:maps}

\subsection{Maps and masks}

The unWISE Blue overdensity is constructed from the number of galaxies $N_i$ in pixel $i$, corrected for stellar density, imaging depth, and the observed fraction of the pixel:
\begin{equation}
n_i=\frac{N_i\,w_i^{\rm star}w_i^{\rm depth}}{f_i^{\rm area}\,\Omega_{\rm pix}},\qquad
g_i=\frac{n_i}{\bar n_g}-1.
\label{eq:overdensity}
\end{equation}
Here $f_i^{\rm area}\Omega_{\rm pix}$ is the effective unmasked area and $w_i^{\rm star}$ and $w_i^{\rm depth}$ are the correction weights. We use the $N_{\rm side}=128$ Blue-sample weight maps of Ref.~\cite{Krolewski_2022}, also adopted in Ref.~\cite{Farren_2024}, upgraded to $N_{\rm side}=2048$ without interpolation. The weights are inverse predicted densities from fits to the \textit{Gaia} stellar density (linear) and the W2 $5\sigma$ limiting magnitude (three-parameter piecewise linear). The galaxy mask is the unWISE mask of Refs.~\cite{Krolewski_2020,Farren_2024}: the \textit{Planck} 2018 lensing mask, smoothed with a $1^\circ$ Gaussian, with additional cuts around bright stars and nearby galaxies. Its mean weight is 0.586. We write the galaxy mask as $\mathcal M_g$, with weight $\mathcal M_{g,i}$ in pixel $i$, so that the mean surface density per steradian is
\[
\bar n_g=\frac{\sum_i \mathcal M_{g,i} N_iw_i^{\rm star}w_i^{\rm depth}}
{\sum_i \mathcal M_{g,i} f_i^{\rm area}\Omega_{\rm pix}},
\]
with both sums restricted to unmasked pixels of nonzero effective area.

The temperature fields are the \textit{Planck} PR3 full-mission HFI intensity maps at 100, 143, 217, and 353~GHz, using the full detector-set intensity product at 353~GHz. We divide each by $\Tcmb=2.7255$~K and apply the temperature mask $\mathcal M_\nu$, the product of the GAL060 Galactic mask, a frequency-specific point-source mask, and a valid-pixel mask. All mask edges, including those of the polarization and alternative-mask tests, are apodized with the \textsc{NaMaster} C2 window over $1^\circ$. The mean weights of the galaxy--temperature mask products are 0.480, 0.477, 0.468, and 0.457, in frequency order. We subtract each map's mean over its mask: from $g$ before filtering, from $g^2$ after squaring, and from temperature before measuring the spectra. The simulations undergo the same processing.

\subsection{Filter, bandpowers, and analysis ranges}
\label{sec:ranges}

Before squaring the galaxy overdensity, we remove its modes with $\ell<20$ and $\ell>3000$, following Ref.~\cite{Farren_2023}, so the filter is $w_\ell=1$ for $20\leq\ell\leq3000$ and zero otherwise. Throughout, $g^2$ denotes the square of this filtered field. The lower cut suppresses large-scale systematics, while the upper cut limits the contribution from small scales. Galaxy multipoles above the measured temperature range still contribute to $g^2T$, so the model applies the same filter to both galaxy legs (\cref{eq:compression_h}). We measure 24 bandpowers over $100\leq L<2500$, identically for data and simulations. The analysis ranges of \cref{sec:estimator} retain 14, 14, 22, and 22 bandpowers at 100, 143, 217, and 353~GHz, respectively.

The upper limits follow from the per-bandpower $g^2T$ S/N against the null covariance. At 100~GHz the S/N falls below unity in most bins above $L\simeq1800$, so we cut at $L=2000$. We use the same limit at 143~GHz so that the two tSZ-dominated channels share one range, although the 143~GHz S/N stays above unity up to the last measured bin, centered at $L=2450$. At 217 and 353~GHz the S/N exceeds 2.6 in every bin, so we keep all bins below $L=2500$.

The lower limits guard against Galactic foregrounds and exclude scales that add little signal. At 217 and 353~GHz, Galactic dust emission rises toward large scales, and the lowest retained 353~GHz bandpower, $300\leq L<400$, already changes by 41\% when Galactic sky is added to the mask (\cref{sec:tests}). We therefore exclude $L<300$ in these channels. At 100 and 143~GHz, the tSZ signal is measured against the primary CMB, whose power rises toward large scales, so the bandpowers below $L=600$ add little S/N. Extending these channels down to $L=400$ raises the $g^2T$ S/N by less than 2\%. \Cref{sec:robustness} shows how removing the $300\leq L<400$ bandpower at 353~GHz, or adding the $400\leq L<600$ bandpowers at 100 and 143~GHz, changes the pressure constraints.

\subsection{Shot noise}
\label{sec:shotnoise}
Squaring the galaxy field pairs each galaxy with itself. This introduces into $g^2T$ the standard Poisson shot noise of galaxy polyspectra \cite{Lacasa_2014,Harscouet_2026}, which we adapt here to the filter and catalogue weights. We write the overdensity of \cref{eq:overdensity} as a sum over the catalogue galaxies $j$, placed at the centers $\hat{\bm n}_j$ of their assigned pixels,
\begin{equation}
 g(\hat{\bm n})=\sum_j\omega_j\,\delta_{\rm D}(\hat{\bm n}-\hat{\bm n}_j)-1,\qquad
 \omega_j=\frac{w_i}{f_i^{\rm area}\,\bar n_g},
 \label{eq:catalogue_sum}
\end{equation}
where $\delta_{\rm D}$ is the Dirac delta function on the sphere, $\bar n_g$ is the surface density per steradian, $i$ is the pixel containing galaxy $j$, and $w_i=w_i^{\rm star}w_i^{\rm depth}$. The galaxy filter $w_\ell$ acts in real space as the kernel $\tilde w(\mu)=\sum_\ell(2\ell+1)\,w_\ell P_\ell(\mu)/4\pi$, where $\mu$ is the cosine of the angular separation and $P_\ell$ is a Legendre polynomial. Filtering therefore removes the monopole and replaces each delta function in \cref{eq:catalogue_sum} by the kernel $\tilde w$ centered on that galaxy, so the filtered $g^2$ is a double sum over galaxy pairs $(j,k)$. The terms with $k=j$ form the shot-noise map
\begin{equation}
 g^2_{\rm shot}(\hat{\bm n})=\sum_j\omega_j^2\,\tilde w^2(\hat{\bm n}\cdot\hat{\bm n}_j)
 =\frac{1}{\bar n_g}\int\dd\Omega'\,\tilde w^2(\hat{\bm n}\cdot\hat{\bm n}')\,g_{\omega^2}(\hat{\bm n}'),\qquad
 g_{\omega^2}(\hat{\bm n})=\bar n_g\sum_j\omega_j^2\,\delta_{\rm D}(\hat{\bm n}-\hat{\bm n}_j).
 \label{eq:self_pair_map}
\end{equation}
Thus $g^2_{\rm shot}$ is the convolution of $g_{\omega^2}$ with the kernel $\tilde w^2$, divided by $\bar n_g$. On the full sky, this kernel depends only on the angular separation, so, as for smoothing by an instrumental beam, the convolution multiplies the spherical-harmonic coefficients by a factor that depends only on $L$. Cross-correlating $g^2_{\rm shot}$ with $X$, using the estimator of \cref{eq:skew}, gives the shot-noise contribution to the skewed bispectrum,
\begin{equation}
\begin{aligned}
 C_L^{g^2X,\,\mathrm{shot}}
 &=\frac{1}{2L+1}\sum_M a_{LM}^{g^2_{\rm shot}}a_{LM}^{X*}\\
 &=\frac{C_L^{g_{\omega^2}X}}{\bar n_g}\sum_{\ell_1\ell_2}\frac{(2\ell_1+1)(2\ell_2+1)}{4\pi}
 \begin{pmatrix}\ell_1&\ell_2&L\\0&0&0\end{pmatrix}^{\!2}w_{\ell_1}w_{\ell_2}.
\end{aligned}
 \label{eq:shot_uniform}
\end{equation}
For uniform catalogue weights and full pixel areas, $\omega_j=1/\bar n_g$ and $g_{\omega^2}=1+g$, so \cref{eq:shot_uniform} reduces to the standard Poisson shot noise, with $C_L^{gX}$ in place of $C_L^{g_{\omega^2}X}$ \cite{Lacasa_2014,Harscouet_2026}. The shot noise measured on a Poisson mock catalogue with uniform catalogue weights agrees with this expression to 1.7\%.

Because the mask couples multipoles, we construct $g^2_{\rm shot}$ before applying it. In pixel $i$, $g_{\omega^2}=N_i(w_i/f_i^{\rm area})^2/(\Omega_{\rm pix}\bar n_g)$. We convolve this field with $\tilde w^2/\bar n_g$ as in \cref{eq:self_pair_map}, then apply the galaxy mask and subtract the mean over the mask, as for $g^2$. We cross-correlate the result with $T_\nu$ using the same masks and bandpower estimator as for $g^2T$.

Unlike the shot noise of an auto-spectrum, this term vanishes on average when galaxies and temperature are uncorrelated, so it is part of the signal and we keep it in the detection tests. For the fits, we subtract its measured bandpowers from the data and from each simulation (\cref{eq:paired_cov}). The halo model therefore needs no term for a galaxy paired with itself (\cref{eq:coincidences}).

\subsection{Simulations and foreground variance}
\label{sec:sims}

Both the null covariance and the mock covariance use 300 microwave realizations per channel. Each is the sum of an FFP10 CMB-plus-noise realization \cite{2020_Planck_HFImaps} and a Gaussian foreground map that is independent of the galaxies and adds foreground fluctuations to the covariance. We obtain the foreground spectra by subtracting the mean FFP10 spectra from the measured temperature auto- and cross-spectra and smoothing the difference over $\Delta L=100$. We draw the foreground maps jointly for the four channels, so they are correlated between frequencies as in the data. Averaged over realizations, the total simulated auto-spectra then match the measured ones to within 5\% over $300\leq L<2500$. The foregrounds make up only 1--2\% of the $g^2T$ variance at 100 and 143~GHz, but 7\% at 217~GHz and 59\% at 353~GHz. Compared with CMB-plus-noise simulations alone, they lower the 353~GHz detection significances (\cref{sec:detection}) from $39.0\sigma$ to $35.5\sigma$ for $gT$ and from $31.8\sigma$ to $28.0\sigma$ for $g^2T$, while for the rest of the channels the change is at most $1.1\sigma$.

For fixed galaxy maps, both estimators are linear in temperature, so only the two-point statistics of the foregrounds enter the covariance. The simulated foregrounds, however, have the same expected variance everywhere, whereas real foregrounds vary across the sky.  The simulations would then underestimate the $g^2T$ errors if stronger foreground fluctuations occur where the $g^2$ fluctuations are also stronger. We check this at 217 and 353~GHz, where foregrounds contribute most to the errors.

We trace the foreground fluctuations with the channel difference $T_\nu-T_{143}$. In CMB temperature units the CMB has the same amplitude in both channels, so the subtraction largely cancels it. Foreground emission changes with frequency and therefore remains in the difference, together with instrumental noise and a small CMB residual from the different beams.

We apply the common mask $\mathcal M_\nu\mathcal M_{143}$, filter the difference map to $300\leq L<2500$, and divide the sky into patches about $3.7^\circ$ across. The mean squared value of the map in each patch $p$ gives its local variance. We remove the noise and the CMB residual by subtracting the same quantity measured on FFP10 CMB-plus-noise simulations, which share the beams and noise of the data. The remainder is the foreground variance $b_p$. We measure the variance $a_p$ of the masked $g^2$ map with the same filter and patches.

The foreground contribution to the $g^2T$ variance is then approximately proportional to $\sum_p a_pb_p$. We compare it with the contribution of a foreground whose variance is uniform before masking and has the same total power. Its variance in patch $p$ is then proportional to the average $m_p$ of the squared mask $(\mathcal M_\nu\mathcal M_{143})^2$ within the patch, and the ratio of the two contributions is
\begin{equation}
 \Phi=\frac{\sum_p a_pb_p\,\sum_p m_p}{\sum_p a_pm_p\,\sum_p b_p}.
\end{equation}
Thus $\Phi=1$ if sky variations of the foreground leave its contribution to the $g^2T$ variance unchanged, and $\Phi>1$ if the simulations underestimate it.

After accounting for the masks, the foreground and $g^2$ variances are nearly uncorrelated between patches ($|r|<0.08$). We find $\Phi=0.94$--1.02, also when $300\leq L<1000$ and $1000\leq L<2500$ are tested separately, so spatial variations change the foreground contribution to the variance by at most 6\%. As a separate check, drawing the foregrounds from a smooth fitted model of the temperature spectra changes the bandpower errors by at most 1\%. These tests support the foreground treatment used in the simulations.

\section{Detection and measurement tests}
\label{sec:detection}

\subsection{Signal-to-noise ratio and detection significance}

Using the null covariance $\widehat{\sf C}_0$ defined in
\cref{sec:detect}, let $\bm d$ contain the $N_{\rm bin}$
measured bandpowers after subtracting the mean simulated
spectrum, and define
$Q=\bm d^{\mathsf T}\widehat{\sf C}_0^{-1}\bm d$.
The Hartlap-corrected signal-to-noise ratio is
\cite{Hartlap_2006}
\begin{equation}
 {\rm S/N}=(h_{\rm H}Q)^{1/2},\qquad
 h_{\rm H}=\frac{N_{\rm sim}-N_{\rm bin}-2}{N_{\rm sim}-1},
 \label{eq:snr}
\end{equation}
where $N_{\rm sim}=300$, giving $h_{\rm H}=0.7559$ for the
joint measurement with $N_{\rm bin}=72$.

\begin{figure*}[t]
\includegraphics[width=0.9\textwidth]{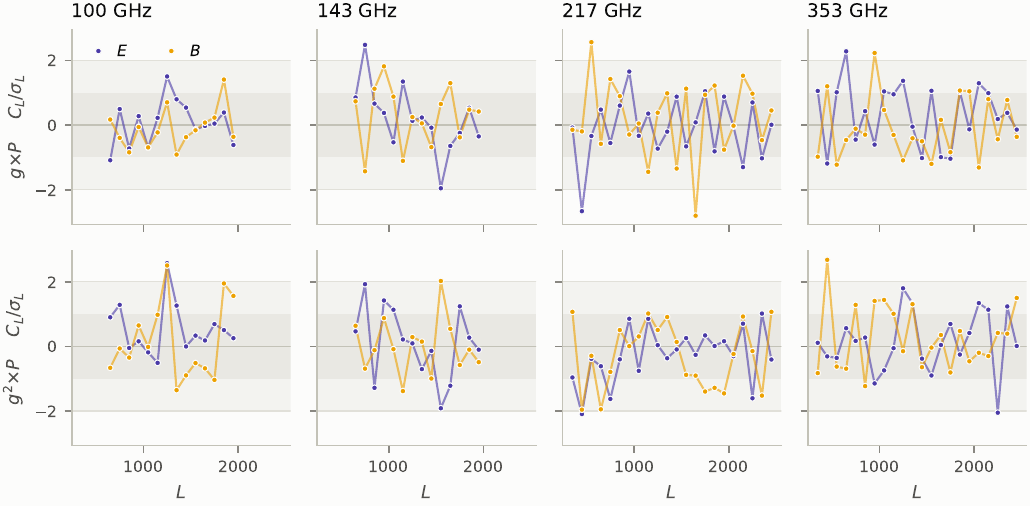}
\caption{Polarization null spectra for $g\times E_\nu$ and $g\times B_\nu$ (top) and $g^2\times E_\nu$ and $g^2\times B_\nu$ (bottom), at each frequency (columns). Each bandpower is the null spectrum itself, shown in units of its statistical uncertainty, estimated from 300 FFP10 polarization realizations cross-correlated with the observed galaxy maps. The shaded bands mark deviations of one and two standard deviations from zero. Only the analysis ranges are shown. PTEs are listed in \cref{tab:null_summary}.}
\label{fig:null}
\end{figure*}

To calculate the detection significance, we need the probability that uncorrelated galaxy and temperature fields would produce a value of $Q$ at least as large as measured. Under this null
hypothesis, we assume that the observed and simulated bandpower vectors are independent draws from the same Gaussian distribution. Accounting for the uncertainty in the mean and covariance
estimated from $N_{\rm sim}$ simulations gives
\cite{Hotelling_1931,Sellentin_2016}
\begin{equation}
 \frac{N_{\rm sim}-N_{\rm bin}}
 {N_{\rm bin}(N_{\rm sim}-1)}
 \frac{N_{\rm sim}}{N_{\rm sim}+1}\,Q
 \sim F_{N_{\rm bin},N_{\rm sim}-N_{\rm bin}},
 \label{eq:hotelling}
\end{equation}
where $F$ is the Fisher distribution and the factor $N_{\rm sim}/(N_{\rm sim}+1)$ accounts for estimating the subtracted mean from the simulations. This $F$ distribution gives the probability of exceeding the measured $Q$, which we express as the one-sided Gaussian significance reported in \cref{tab:summary}. Thus S/N and the detection significance are two summaries of the same statistic $Q$, its size and its probability if the fields are uncorrelated.

\subsection{Polarization and mask tests}
\label{sec:tests}

We assess possible contamination using polarization null tests and changes to the sky masks. We estimate the polarization-null covariance from 300 FFP10 polarization realizations per channel, cross-correlated with the observed galaxy fields. The correlations are consistent with zero also in the lowest fitted bandpower at 217 and 353~GHz, where the temperature measurements are most sensitive to the mask (\cref{fig:null,tab:null_summary}).

\begin{figure*}[t]
\includegraphics[width=0.95\textwidth]{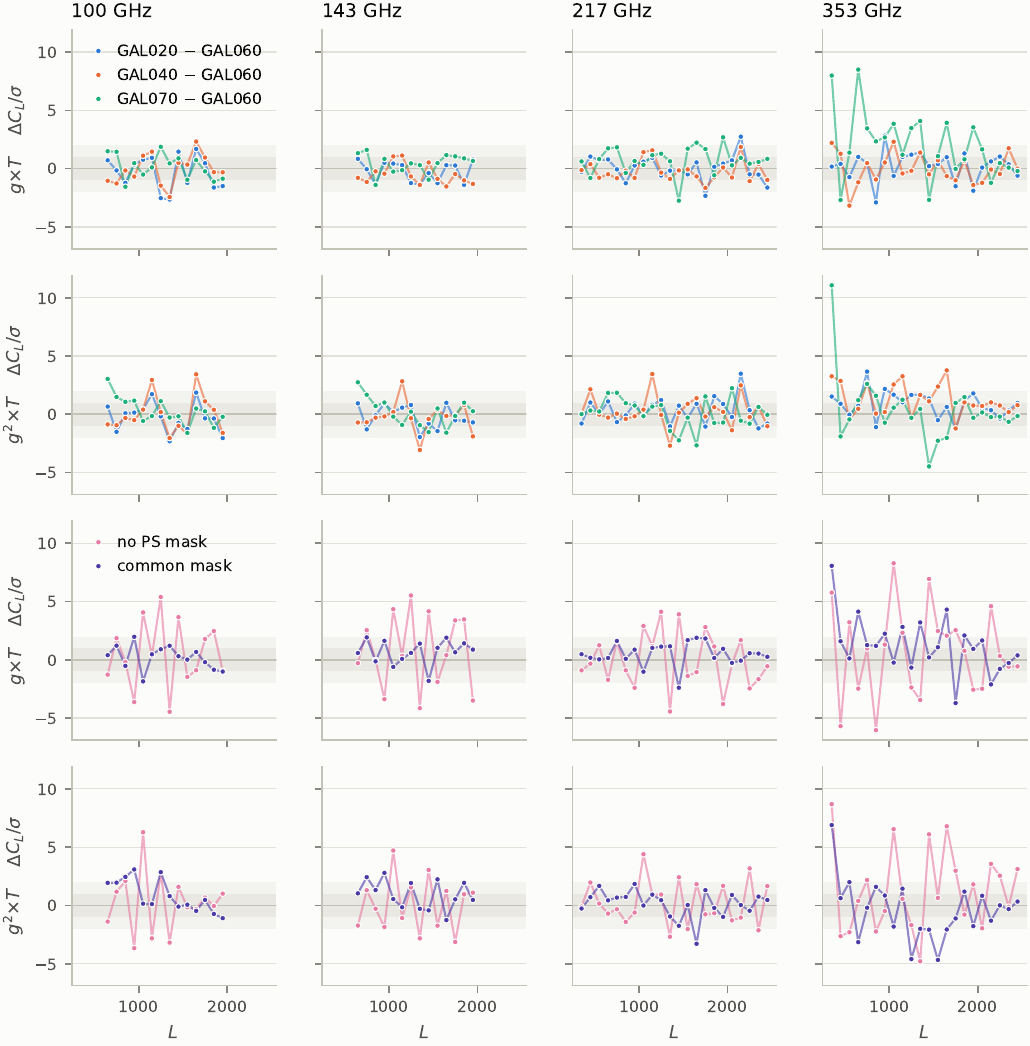}
\caption{Changes in $g\times T_\nu$ and $g^2\times T_\nu$ relative to the fiducial measurements. The top two rows vary the Galactic mask while keeping the point-source masks fixed. In each of the bottom two rows, the magenta curves (``no PS mask'') remove the point-source masks while retaining GAL060; the purple curves (``common mask'') replace both the Galactic and point-source masks with the \textit{Planck} common intensity mask. Rows one and three show $g\times T_\nu$; rows two and four show $g^2\times T_\nu$. Each difference is shown in units of its uncertainty, estimated by applying both mask choices to the same 300 FFP10 CMB-plus-noise realizations. Only the analysis ranges are shown. Horizontal bands mark deviations of one and two standard deviations from zero. \Cref{tab:null_summary} gives the combined significance of the changes across all multipole bins in the analysis ranges.}
\label{fig:masks}
\end{figure*}

We vary the sky masks to assess sensitivity to Galactic
emission and bright sources. For each change, we measure the
overall amplitude shift and the significance of the differences
across multipole bins. The amplitude $A$ relates the
alternative-mask spectrum $\bm d'$ to the fiducial spectrum
$\bm d$:
\begin{equation}
 A=\frac{\bm d^{\mathsf T}\widehat{\sf C}_0^{-1}\bm d'}
 {\bm d^{\mathsf T}\widehat{\sf C}_0^{-1}\bm d}.
 \label{eq:amplitude}
\end{equation}
The percentage change $100(A-1)$ is reported for $g^2T$ in
\cref{tab:summary}. We estimate the covariance of the differences
across bins by applying both masks to the same 300 FFP10
CMB-plus-noise realizations, and calculate their significance
using \cref{eq:hotelling}
(\cref{fig:masks,tab:null_summary}). These errors include CMB
and instrumental-noise fluctuations, but not foreground or
galaxy sample variance.

\begin{table*}[t]
\centering
\setlength{\tabcolsep}{1.5pt}
\small
\renewcommand{\arraystretch}{1.2}
\begin{tabular}{l|cccc|cccc}
\hline
& \multicolumn{4}{c|}{$g\times X_\nu$} & \multicolumn{4}{c}{$g^2\times X_\nu$} \\
Test & 100 & 143 & 217 & 353 & 100 & 143 & 217 & 353 \\
\hline
\multicolumn{9}{l}{\textit{Polarization nulls ($X=E,B$)}} \\
$E$-mode null, PTE & 0.97 & 0.51 & 0.49 & 0.47 & 0.58 & 0.25 & 0.79 & 0.84 \\
$B$-mode null, PTE & 0.98 & 0.52 & 0.07 & 0.80 & 0.18 & 0.79 & 0.31 & 0.41 \\
$E$/$B$ null, $300\le L<400$, PTE & -- & -- & 0.94/0.89 & 0.29/0.33 & -- & -- & 0.34/0.29 & 0.91/0.41 \\
\hline
\multicolumn{9}{l}{\textit{Mask comparisons ($X=T$)}} \\
GAL020 $-$ GAL060 [$\sigma$] & 2.0 & $<1$ & $<1$ & 1.5 & 1.0 & $<1$ & $<1$ & 2.3 \\
GAL040 $-$ GAL060 [$\sigma$] & 1.2 & $<1$ & $<1$ & 1.6 & 2.8 & 1.3 & 2.1 & 4.3 \\
GAL070 $-$ GAL060 [$\sigma$] & $<1$ & $<1$ & 2.0 & 10.7 & 1.1 & 1.2 & 1.8 & 9.0 \\
\hline
Common mask $-$ fiducial [$\sigma$] & $<1$ & 1.2 & $<1$ & 8.4 & 2.7 & 2.4 & $<1$ & 7.5 \\
No PS mask $-$ fiducial [$\sigma$] & 6.4 & 7.8 & 6.3 & 12.6 & 6.8 & 5.6 & 3.4 & 11.3 \\
\hline
\end{tabular}
\caption{Measurement tests over the analysis ranges, with frequencies in GHz. Polarization rows give PTEs for consistency with zero, with paired entries ordered as $E/B$. Mask rows give the significance, in units of $\sigma$, of changes across the multipole bins relative to the fiducial measurement, using errors from paired CMB-plus-noise simulations. Overall $g^2T$
amplitude changes are listed separately in \cref{tab:summary}. PS denotes point sources.}
\label{tab:null_summary}
\end{table*}

We replace GAL060 with GAL020, GAL040, or GAL070 while retaining the frequency-specific point-source masks. We also use the \textit{Planck} common intensity mask \cite{2020_Planck_SMICA}, which changes both Galactic coverage and source exclusions. The resulting differences are shown in \cref{fig:masks,tab:null_summary}. We test the effect of excluding the mask-sensitive lowest 353~GHz bandpower from the fits in \cref{sec:robustness}.

To test sensitivity to bright sources, we remove the
point-source masks while keeping GAL060. This produces
detectable changes across the multipole bins in every channel,
while the overall $g^2T$ amplitudes change by at most 1.2\%.
We retain the frequency-specific masks to exclude these sources.
We also check whether source masking removes appreciable tSZ
signal by repeating the $g^2\times y$ measurement on the NILC
map with the 143 and 353~GHz point-source masks and without
point-source masking. The overall amplitude changes by about
1\%, indicating a small net sensitivity to these mask choices.
This comparison includes other differences between the masks
and residual CIB emission in the NILC map
\cite{2016Planck_ymap}.

\section{Halo model}
\label{sec:model}

The halo model combines the distribution of galaxies within halos with models for dust emission and electron pressure to predict $gT$ and $g^2T$. We adopt a flat $\Lambda$CDM cosmology with the \textit{Planck} 2018 TT,TE,EE+lowE+lensing+BAO parameters \cite{2020Planck_cosmoparams}: $h=0.6766$, $\Omega_bh^2=0.02242$, $\Omega_ch^2=0.11933$, $n_s=0.9665$, $\ln(10^{10}A_s)=3.047$, $\tau_{\rm reio}=0.0561$, and one massive neutrino species with $\sum m_\nu=0.06$~eV, giving $\Omega_m=0.311$ and $\sigma_8=0.810$. CAMB \cite{Lewis_2000} supplies the background expansion and the linear matter power spectrum $P_{\rm lin}$. Halo masses are $M=M_{200c}$, enclosing a mean density 200 times the critical density, in physical $M_\odot$. We use the halo mass function of Ref.~\cite{Tinker_2008} and the linear halo bias of Ref.~\cite{Tinker_2010_bias}, both evaluated with COLOSSUS \cite{Diemer_2018}, and obtain the second-order bias from the peak--background split of the mass-function fit of Ref.~\cite{Tinker_2010_bias}. Halo concentrations follow the relation of Ref.~\cite{Duffy_2008}. Mass integrals run over $10^{10}\leq M/M_\odot\leq10^{16}$; halos above $10^{15.8}\,M_\odot$ contribute less than 0.1\% of the one-halo $g^2T$ tSZ signal (\cref{fig:massweights}).

\subsection{Projection}
\label{sec:compression}

For projected fields $X$ and $Y$, the Limber approximation gives
\begin{equation}
C_L^{XY}=\int \frac{\mathrm{d}\chi}{\chi^2}\,W_X(\chi)W_Y(\chi)
P_{XY}\!\left(\frac{L+1/2}{\chi},z\right),
\label{eq:limber}
\end{equation}
where $\chi$ is comoving distance, $z=z(\chi)$ is the corresponding redshift, $W_X$ and $W_Y$ are projection kernels, and the three-dimensional spectrum $P_{XY}$ contains one- and two-halo terms. For galaxies, $W_g$ is the normalized unWISE redshift distribution, converted from redshift to comoving distance.

To predict the skewed bispectrum, we use the Limber and flat-sky approximations, as in Ref.~\cite{Farren_2023}. \Cref{eq:compression} then becomes
\begin{equation}
C_L^{g^2T_\nu}=\int\frac{\mathrm{d}^2\ell}{(2\pi)^2}\,w_{\ell}\,w_{|\bm L-\bm\ell|}
\int\frac{\mathrm{d}\chi}{\chi^4}\,W_g^2(\chi)\,W_{T_\nu}(\chi)\,
B^{ggT_\nu}(k_1,k_2,k_3;z),
\label{eq:compression_h}
\end{equation}
with $k_1=\ell/\chi$, $k_2=|\bm L-\bm\ell|/\chi$, and $k_3=L/\chi$. The integral runs over all triangles formed by the temperature multipole $\bm L$ and the two filtered galaxy multipoles $\bm\ell$ and $\bm L-\bm\ell$. The halo-model bispectrum is
\cite{Scoccimarro2001,COORAY_2002,Smith_2008}
\begin{equation}
B^{ggT_\nu}=B^{ggT_\nu}_{1h}
           +B^{ggT_\nu}_{2h}
           +B^{ggT_\nu}_{3h},
\label{eq:bisp_halo}
\end{equation}
with
\begin{align}
B^{ggT_\nu}_{1h}&=\big\langle u^g_{k_1}u^g_{k_2}u^{T_\nu}_{k_3}\big\rangle_M,\label{eq:bisp_1h}\\
B^{ggT_\nu}_{2h}&=\big\langle b^{(1)}u^g_{k_1}u^g_{k_2}\big\rangle_M\big\langle b^{(1)}u^{T_\nu}_{k_3}\big\rangle_M P_{\rm lin}(k_3)+\text{2 perm.},\label{eq:bisp_2h}\\
B^{ggT_\nu}_{3h}&=\big\langle b^{(1)}u^g_{k_1}\big\rangle_M\big\langle b^{(1)}u^g_{k_2}\big\rangle_M\big\langle b^{(1)}u^{T_\nu}_{k_3}\big\rangle_M\big[2F_2(\bm k_1,\bm k_2)P_{\rm lin}(k_1)P_{\rm lin}(k_2)+\text{2 perm.}\big]\nonumber\\
&\quad+\big\langle b^{(1)}u^g_{k_1}\big\rangle_M\big\langle b^{(1)}u^g_{k_2}\big\rangle_M\big\langle b^{(2)}u^{T_\nu}_{k_3}\big\rangle_M P_{\rm lin}(k_1)P_{\rm lin}(k_2)+\text{2 perm.},
\end{align}
where $\langle\cdots\rangle_M=\int\mathrm{d}M\,(\mathrm{d}n/\mathrm{d}M)\cdots$ integrates over the halo mass function, $u^X_k$ is the Fourier-space halo weight for field $X$, $b^{(1)}$ and $b^{(2)}$ are the linear and quadratic halo biases, and $F_2$ is the tree-level perturbation-theory kernel. The subscript $M$ distinguishes this mass integral from expectations over the galaxy occupation at fixed halo mass, such as $\langle N_c\rangle$ (\cref{sec:profiles}). For galaxies, $u^g_k$ is the Fourier transform of the galaxy number profile of the halo (\cref{sec:profiles}) divided by the mean comoving galaxy number density $n_g(z)$, which converts galaxy counts into overdensity. This density is per unit comoving volume, unlike the surface density $\bar n_g$ of \cref{sec:maps}. For the temperature, $u^{T_\nu}_k$ is the Fourier transform of the tSZ or CIB temperature profile of the halo. All predictions include the \textit{Planck} bandpass unit conversions and color corrections \cite{2014_Planck_SpectralResponse,2014_CIB_colorcorr}.

\subsection{Galaxy occupation and emission profiles}
\label{sec:profiles}

\paragraph{Galaxy occupation.} The occupation consists of a Bernoulli central galaxy and Poisson satellites conditional on the central being present:
\begin{equation}
 \langle N_c\rangle=\frac12\left[1+\operatorname{erf}\left(
 \frac{\log_{10}(M/M_{\rm min}^{\rm gal})}{\sigma_{\log M}}\right)\right],
 \qquad \langle N_s\rangle=\langle N_c\rangle
 \left[\frac{\max(0,M-M_{\rm min}^{\rm gal})}{M_1'}\right]^{\alpha_s}.
\end{equation}
Central galaxies lie at halo centers. Satellites follow an NFW profile with the scale radius of their host halo, truncated at $r_{\rm max}=\lambda_{\rm sat}R_{200c}$, where $\lambda_{\rm sat}$ is a free parameter \cite{Kusiak_2022}. We write $u_s(k|M,z)$ for the Fourier transform of this profile, normalized to unity as $k\to0$. In the one-halo term, $gT$ weights each halo by its mean number of galaxies, whereas $g^2T$ weights it by its mean number of distinct pairs, through $\langle N_cN_s\rangle$ and $\langle N_s(N_s-1)\rangle$ \cite{COORAY_2002}. Because the number of pairs grows faster with halo mass than the number of galaxies, $g^2T$ is more sensitive to the occupation of massive halos.

We fit the five occupation parameters to 19 galaxy
auto-spectrum bandpowers over $100\leq L<2000$, including
magnification and a Poisson shot-noise level fixed by the
galaxy weights. We use a Gaussian likelihood with a fixed
analytic covariance,
\begin{equation}
 \operatorname{Cov}\big(\hat C_L^{gg},\hat C_{L'}^{gg}\big)={\sf C}_{\rm G}
             +{\sf C}_{\rm 1h}
             +{\sf C}_{\rm shot}.
\end{equation}
The Gaussian term ${\sf C}_{\rm G}$ is computed from the
measured $C_L^{gg}$, including shot noise, with mask coupling
computed using \textsc{NaMaster}.

\begin{figure*}[t]
\includegraphics[width=0.82\textwidth]{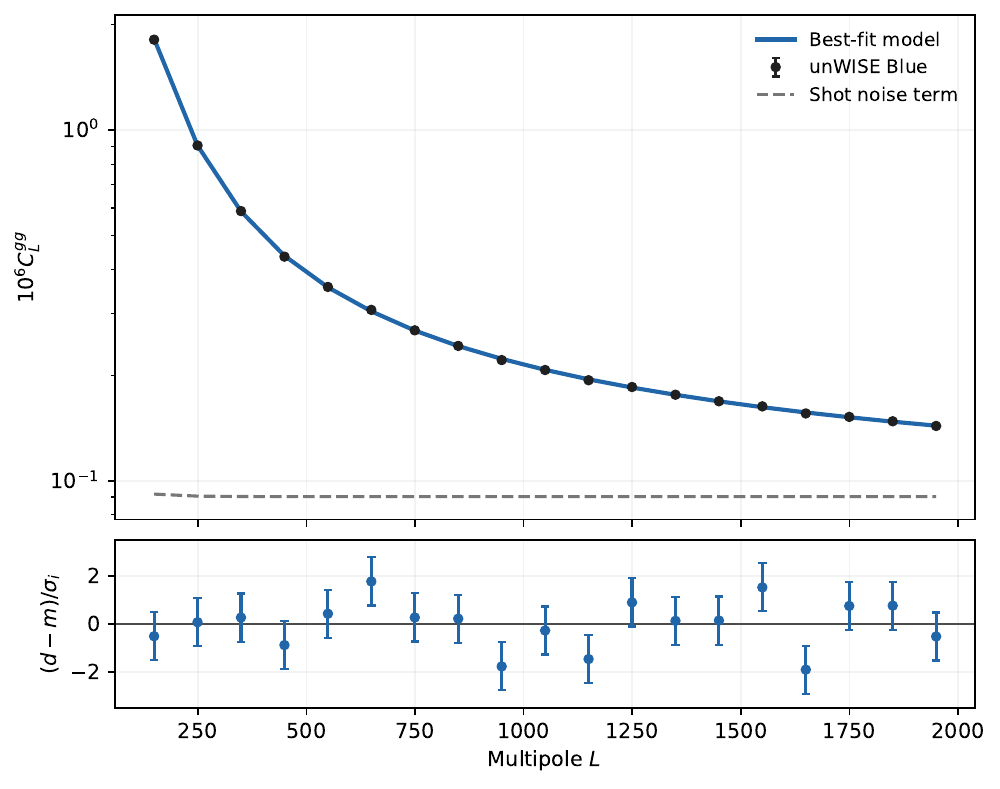}
\caption{Measured unWISE Blue galaxy auto-spectrum and the five-parameter HOD fit over $100\leq L<2000$. The best-fit model (solid) includes magnification and the fixed shot-noise term (dashed). The lower panel shows residuals divided by the diagonal analytic errors, while the fit uses the full covariance.}
\label{fig:cggfit}
\end{figure*}

For ${\sf C}_{\rm 1h}$, we integrate the one-halo trispectrum
of four distinct galaxies over halo mass and redshift.
The occupation moments count configurations containing four
satellites or one central and three satellites, weighted by
their spatial profiles. We evaluate these moments using the
unWISE Blue HOD parameters of Ref.~\cite{Kusiak_2023}, held
fixed for the covariance while the occupation parameters of
the mean spectrum are fitted. The connected shot-noise term ${\sf C}_{\rm shot}$ contains the contribution in which the
same galaxy appears four times. We calculate it from the
observed counts and fourth powers of the galaxy weights,
including the effective pixel areas and mask weights.
Both connected terms are projected into bandpowers, with
the survey geometry approximated by an effective sky area. We neglect connected terms involving two or three distinct
galaxies, multi-halo trispectra, and super-sample covariance
from density fluctuations on scales larger than the survey. The covariance is analytic, so no Hartlap correction is needed.
The best fit gives $\chi^2_{\rm red}=18.85/(19-5)=1.35$
(${\rm PTE}=0.17$; \cref{fig:cggfit,tab:cggparams}).

The best fit reaches the upper prior bounds of two parameters, $M_{\rm min}^{\rm gal}=10^{13.5}\,M_\odot$ and $\lambda_{\rm sat}=3$. To test whether these bounds restrict the fit, we extend them to $10^{14}\,M_\odot$ and 5. The fit improves by only $\Delta\chi^2=0.98$. Fixing $\lambda_{\rm sat}$ at 3 or 5, and re-optimizing the other parameters, changes $\chi^2$ by only 0.16, and fixing $\log_{10}(M_{\rm min}^{\rm gal}/M_\odot)$ at 13 or 14 changes it by 0.11. The galaxy auto-spectrum therefore constrains these two parameters only weakly, and quite different occupations fit it almost equally well.

The large-scale bias of the adopted best fit agrees with previous measurements for the unWISE Blue sample. The effective linear galaxy bias is 1.47 at $z=0.6$ and 1.50 averaged over the redshift distribution. For comparison, Ref.~\cite{Kusiak_2022} finds $1.50^{+0.01}_{-0.02}$ from the galaxy auto-spectrum and its cross-correlation with \textit{Planck} CMB lensing, and cross-correlations with spectroscopic samples give 1.52 at $z=0.6$ \cite{Krolewski_2020}. The satellite fraction at $z=0.6$, 0.07, is lower than the $0.23^{+0.06}_{-0.07}$ of Ref.~\cite{Kusiak_2022} and than the fractions from cross-correlations with BOSS CMASS galaxies, which fall from about 20\% to 10\% over $0.45<z<0.75$ \cite{Krolewski_2025}. Because the satellite fraction averages over halos of all masses, it does not by itself set the number of galaxy pairs in the massive halos that dominate the $g^2T$ tSZ signal.

Throughout the analysis, we fix the occupation to the adopted best fit in \cref{tab:cggparams}, so the quoted uncertainties do not include the uncertainty of the galaxy occupation. This matters most for $g^2T$, because an occupation with more galaxy pairs in massive halos can produce the same $g^2T$ signal with less pressure. Propagating this uncertainty would require varying the occupation together with the emission parameters. The correlations with temperature also omit magnification, which we include in the galaxy auto-spectrum fit. \Cref{sec:robustness} estimates its contribution.

\begin{table}[t]
\centering
\renewcommand{\arraystretch}{1.2}
\begin{tabular}{c|c|c}
\hline
Parameter & Flat prior & Best fit \\
\hline
$\alpha_s$ & $[0.1,3.0]$ & $1.51$ \\
$\sigma_{\log M}$ & $[0.02,1.5]$ & $1.17$ \\
$\log_{10}(M_{\rm min}^{\rm gal}/M_\odot)$ & $[10.9,13.5]$ & $13.50$ \\
$\log_{10}(M_1'/M_\odot)$ & $[11.5,15.0]$ & $13.97$ \\
$\lambda_{\rm sat}$ & $[0.2,3.0]$ & $3.00$ \\
\hline
$\chi^2_{\rm red}$ & & $18.85/14=1.35$ \\
PTE & & 0.17 \\
Satellite fraction at $z=0.6$ & & 0.07 \\
\hline
\end{tabular}
\caption{Priors and best-fit parameters of the
five-parameter HOD model for the galaxy auto-spectrum.
The best fit is determined within the stated prior bounds and held fixed in all microwave fits.
Masses are physical $M_\odot$.}
\label{tab:cggparams}
\end{table}

\paragraph{CIB emission.} We describe the CIB emissivity with the star-formation-efficiency model of Refs.~\cite{Maniyar_2021,Yan_2024},
\begin{equation}
 \eta(M,z)=\eta_{\rm max}\exp\left[-\frac{\ln^2(M/M_{\rm eff})}{2\,[\sigma_M^{\rm CIB}(M,z)]^2}\right],
 \qquad \mathrm{SFR}(M,z)=\eta(M,z)\frac{\Omega_b}{\Omega_m}\dot M(M,z),
 \label{eq:cib_efficiency}
\end{equation}
where $M_{\rm eff}$ is the mass at which star formation is most efficient, and $\eta_{\rm max}$ is the fraction of accreted baryons converted into stars at that mass. We keep $M_{\rm eff}$ independent of redshift. The width in natural logarithmic mass is
\begin{equation}
 \sigma_M^{\rm CIB}(M,z)=\begin{cases}
 \sigma_{M,0}^{\rm CIB},&M<M_{\rm eff},\\
 \sigma_{M,0}^{\rm CIB}\left[1-\dfrac{\tau}{z_c}\max(0,z_c-z)\right],&M\geq M_{\rm eff},
 \end{cases}
 \label{eq:cib_width}
\end{equation}
with $\sigma_{M,0}^{\rm CIB}=2.74$, $\tau=0.5$, and $z_c=2.15$. These are the posterior means of Ref.~\cite{Yan_2024} for this model, which that work calls M21.

We multiply the central emission by an infrared occupation probability, described by an error-function transition at $M_{\rm min}^{\rm CIB}=10^{11.38}\,M_\odot$ \cite{Yan_2024}, with width 0.302 in $\log_{10}M$. Ref.~\cite{Yan_2024} allows this threshold to evolve as $\log_{10}(M_{\rm min}^{\rm CIB}/M_\odot)=11.38+2.6\,z/(1+z)$, with a broad posterior on the evolution. We instead retain its $z=0$ value at all redshifts.

We evaluate the central accretion rate using the full host mass, without subtracting a fixed mass fraction assigned to subhalos. We integrate satellite emission over the subhalo population following Ref.~\cite{Maniyar_2021}. The effective infrared spectral energy distributions of Ref.~\cite{Bethermin_2013}, integrated over the \textit{Planck} bandpasses as in Ref.~\cite{Maniyar_2021}, convert the star-formation rates into emissivities in the four frequency channels. We calculate the CIB terms by projecting from the emissivity model above using \cref{eq:limber,eq:compression_h}.

Both central and satellite galaxies can supply the self-dust emission introduced in \cref{sec:halomodel}. We construct our model using emission from centrals. Following the $gT$ treatment of Ref.~\cite{Yan_2024}, we separate the central--central contribution from the one-halo galaxy--CIB spectrum and fit its amplitude independently of the remaining CIB emission. Since the $gT$ self-dust contribution is constant for either type of galaxy, this free amplitude also absorbs satellite self-dust. We extend the central-emission model to $g^2T$, where the emitting central belongs to a distinct galaxy pair, and fit its amplitude independently. We neglect satellite self-dust in $g^2T$.

Let $T_{\nu,c}$ denote the dust emission of selected central galaxies, with halo weight $u^{T_{\nu,c}}(M,z)$. We treat this emission as a point source at the halo center, so the weight is independent of the mode $k$. At fixed redshift, for uniform catalogue weights and before angular projection and instrumental smoothing, the total catalogue bispectrum is
\begin{equation}
 B_{\rm tot}^{ggT_\nu}(k_1,k_2,k_3)=B^{ggT_\nu}(k_1,k_2,k_3)
 +\underbrace{\frac{1}{n_g}P_{\rm tot}^{gT_\nu}(k_3)}_{B_{\rm shot}^{ggT_\nu}}
 +\underbrace{\frac{1}{n_g}\Big[P^{gT_{\nu,c}}(k_1)+P^{gT_{\nu,c}}(k_2)\Big]}_{B_{\rm self}^{ggT_\nu}}.
 \label{eq:coincidences}
\end{equation}
The first term is the halo-model bispectrum of \cref{eq:bisp_halo}. Its projection gives the tSZ and CIB terms of \cref{eq:cib_coincident}. The second term contains all pairs in which the same galaxy occupies both galaxy legs. Because $P_{\rm tot}^{gT_\nu}$ includes that galaxy's own emission, it also contains the case in which the emitter and both galaxy entries are the same object. We measure and subtract this entire term through its projection $C_L^{g^2T_\nu,\,\mathrm{shot}}$ (\cref{eq:shot_uniform}).

The final term gives the self-dust contribution that remains when the emitter is one member of a distinct pair. Here $P^{gT_{\nu,c}}$ correlates a galaxy with dust emitted by a different central. Excluding each central's correlation with its own dust emission avoids counting again a contribution already removed by the shot-noise subtraction. Under our assumption that the dust emitter is a selected central galaxy, this excluded correlation is the scale-independent power spectrum
\begin{equation}
 P_{\rm self}^{gT_{\nu,c}}(z)=\frac{1}{n_g}\big\langle\langle N_c\rangle\,u^{T_{\nu,c}}\big\rangle_M.
 \label{eq:own_dust_power}
\end{equation}

The surviving self-dust bispectrum has one- and two-halo parts, $B_{\rm self}^{ggT_\nu}=B_{1h,\rm self}^{ggT_\nu}+B_{2h,\rm self}^{ggT_\nu}$. In the one-halo term, the companion of the emitting central is a satellite in the same halo, giving the occupation factor $\langle N_cN_s\rangle$. In the two-halo term, the companion galaxy lies in a different halo from the emitting central. We calculate the correlation between their halos using the linear matter power spectrum. This gives
\begin{equation}
\begin{aligned}
 B_{1h,\rm self}^{ggT_\nu}&=\frac{1}{n_g^2}\big\langle\langle N_cN_s\rangle\big[u_s(k_1)+u_s(k_2)\big]u^{T_{\nu,c}}\big\rangle_M,\\
 B_{2h,\rm self}^{ggT_\nu}&=\frac{1}{n_g}\big\langle b^{(1)}\langle N_c\rangle\,u^{T_{\nu,c}}\big\rangle_M
 \big[\langle b^{(1)}u^g_{k_1}\rangle_M P_{\rm lin}(k_1)+\langle b^{(1)}u^g_{k_2}\rangle_M P_{\rm lin}(k_2)\big],
\end{aligned}
 \label{eq:own_dust_bispectrum}
\end{equation}
where we suppress mass and redshift arguments for brevity. The two contributions in each expression account for the two choices of which galaxy in the pair supplies the dust. Projecting $B_{\rm self}^{ggT_\nu}$ with \cref{eq:compression_h} and including the same galaxy filters gives $C_L^{g^2T_\nu,\,\mathrm{self}}$.

For $gT$, \cref{eq:own_dust_power} describes the correlation of a central galaxy with its own dust emission. Projecting it with \cref{eq:limber} gives the scale-independent prediction
\begin{equation}
 C^{gT_\nu,\,\mathrm{self}}=\int\frac{\dd\chi}{\chi^2}\,W_gW_{T_\nu}\,P_{\rm self}^{gT_{\nu,c}}.
 \label{eq:own_dust_gt}
\end{equation}
We use this prediction to normalize the self-dust template at 353~GHz,
\begin{equation}
 \mathcal S_L=\frac{C_L^{g^2T_{353},\,\mathrm{self}}}{C^{gT_{353},\,\mathrm{self}}}.
 \label{eq:own_dust_shape}
\end{equation}
Both quantities in this ratio are predictions of the same emission model. This normalization places the independently fitted amplitudes $A_{353}^{gT}$ and $A_{353}^{g^2T}$ on a common scale. The template depends on the sampled $M_{\rm eff}$, while the common factor $\eta_{\rm max}$ cancels. We give the amplitude priors in \cref{tab:g2params} and report their values as $10^{12}A_{353}$.

\paragraph{tSZ emission.} The tSZ signal traces the line-of-sight electron pressure,
\begin{equation}
 y(\hat{\bm n})=\frac{\sigma_T}{m_ec^2}\int P_e\,\dd l_{\rm phys},\qquad
 \left(\frac{\Delta T_\nu}{\Tcmb}\right)_{\rm nr}=f_\nu y,
 \quad f_\nu=x_\nu\coth(x_\nu/2)-4,
 \quad x_\nu=\frac{h\nu}{k_B\Tcmb},
\end{equation}
where $\sigma_T$ is the Thomson cross section, $m_e$ the electron mass, and $f_\nu$ the nonrelativistic response before bandpass averaging. We use the generalized NFW profile \cite{Arnaud_2010,2013Planck_tSZparams_bestfit}
\begin{equation}
 P_e(r|M,z)=P_{500}(M,z)p(x),\qquad
 p(x)=\frac{P_0}{(c_{500}x)^\gamma
 [1+(c_{500}x)^\alpha]^{(\beta_{\rm tSZ}-\gamma)/\alpha}},\quad x=\frac{r}{R_{500}}.
 \label{eq:gnfw}
\end{equation}
$P_0$ sets the profile normalization and $c_{500}=R_{500}/r_s$ its concentration. The inner and outer slopes are $\gamma=0.31$ and $\beta_{\rm tSZ}$, with $\alpha=1.33$ controlling their transition \cite{2013Planck_tSZparams_bestfit}. We convert $M_{200c}$ to true $M_{500c}$ using the concentration relation of Ref.~\cite{Duffy_2008}, then evaluate the pressure at the hydrostatic mass $M_{500}^{\rm HSE}=M_{500c}/B$, with $B=1.25$ \cite{Bolliet_2018,2014_Planck_SZcounts}. The reference radius satisfies $M_{500}^{\rm HSE}=(4\pi/3)500\rho_c(z)R_{500}^3$, with $\rho_c(z)=3H^2(z)/(8\pi G)$. Including the empirical mass exponent, the pressure normalization is \cite{Arnaud_2010}
\begin{equation}
 P_{500}=1.65\times10^{-3}\,h_{70}^2 E(z)^{8/3}
 \left[\frac{h_{70}M_{500}^{\rm HSE}}{3\times10^{14}M_\odot}\right]^{2/3+0.12}
 \,\mathrm{keV\,cm^{-3}},
 \label{eq:p500}
\end{equation}
where $E(z)=H(z)/H_0$ and $h_{70}=H_0/(70\,\mathrm{km\,s^{-1}\,Mpc^{-1}})$. The dimensionless $Y_{5R_{500}}$ (\cref{eq:pressure}) is the volume integral of $p(x)$ within $5R_{500}$ and depends only on this profile. Multiplying it by $P_{500}R_{500}^3$ gives the integrated physical electron pressure. Changing $B$ changes this physical normalization through $P_{500}$ and $R_{500}$ for a halo of fixed true mass, while leaving $Y_{5R_{500}}$ unchanged at fixed $p(x)$.

Massive clusters dominate the $g^2T$ tSZ signal (\cref{sec:massweights}), so we include relativistic corrections for their hot electrons. We assign each halo a temperature using the mass--temperature relation of Ref.~\cite{Arnaud_2005} at $M_{500c}/B$, calculate the SZpack response \cite{Chluba_2012}, and integrate it over the HFI bandpasses \cite{2014_Planck_SpectralResponse}.\footnote{We tabulate responses up to $k_BT=60$~keV and assign the 60-keV response to hotter halos, which occupy only the negligibly abundant upper mass tail at $z\gtrsim3.5$.} The ratio to the nonrelativistic prediction supplies a fixed correction to each bandpower.

\subsection{Host-halo masses}
\label{sec:massweights}

\begin{figure*}[t]
\includegraphics[width=0.9\textwidth]{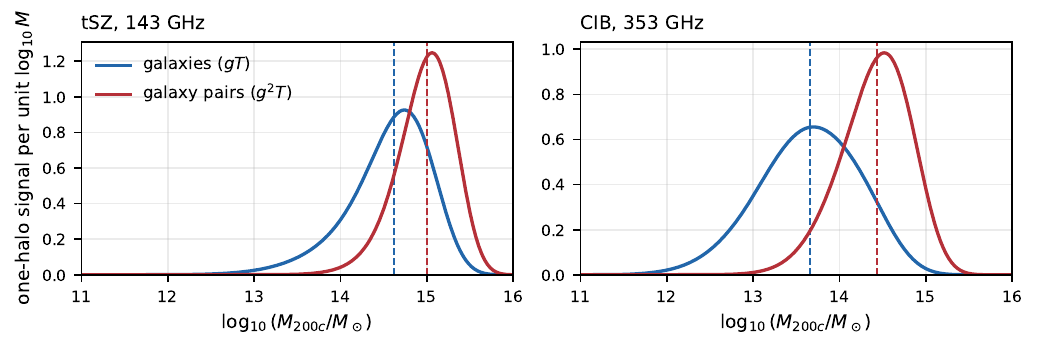}
\caption{One-halo signal distributions over host mass for tSZ at 143~GHz (left) and CIB at 353~GHz (right). Blue curves use the best-fit parameters from the $gT$ fit, and red curves use those from the $g^2T$ fit. Both fits fix the outer pressure slope to $\beta_{\rm tSZ}=4.13$. Curves have unit area, and dashed lines mark the medians. Differences reflect both galaxy weighting and fitted emission parameters. The $g^2T$ CIB curve includes self-dust (\cref{eq:cib_coincident}).}
\label{fig:massweights}
\end{figure*}

To identify which halo masses contribute to the modeled signals, we sum the one-halo contribution over redshift and the fitted bandpowers, using the respective best-fit emission parameters for $gT$ and $g^2T$. We normalize each mass distribution to unit area (\cref{fig:massweights}). For $g^2T$, the one-halo term supplies 87\% of the tSZ signal at 143~GHz, while for the CIB at 353~GHz, it supplies only 39\%, with the remaining signal coming from the two- and three-halo terms.

The adopted galaxy occupation gives a mean host-halo mass of $2.1\times10^{13}\,M_\odot$, comparable to that of Ref.~\cite{Kusiak_2022}. The tSZ signal gives much greater weight to massive halos because the signal per halo grows approximately as $M^{1.8}$, as also occurs for the tSZ power spectrum. The weighting by distinct galaxy pairs further emphasizes massive halos in $g^2T$.

For the one-halo tSZ signal at 143~GHz, the median contributing mass is $10^{15.0}\,M_\odot$ for $g^2T$ and $10^{14.6}\,M_\odot$ for $gT$. Halos above $10^{15}\,M_\odot$ supply 52\% and 18\% of these contributions, respectively. For the one-halo CIB signal at 353~GHz, the corresponding medians are $10^{14.4}$ and $10^{13.7}\,M_\odot$. The concentration of the $g^2T$ tSZ signal in massive clusters explains its sensitivity to fluctuations in their abundance (\cref{sec:clusterterm}).

\section{Covariance and likelihood}
\label{sec:cov}

Below we construct the mock and halo-count terms of \cref{eq:cov_split} and then define the likelihood. For each statistic, ${\sf C}^{\nu\nu'}(L,L')$ denotes the covariance between bandpowers at $L$ and $L'$ in channels $\nu$ and $\nu'$, respectively.

\subsection{Mock covariance}
\label{sec:mockcov}

Because we use the observed galaxy maps, ${\sf C}_{\rm mock}$ retains the full measured $C_L^{g^2g^2}$, including the contribution of the connected galaxy four-point function, which Gaussian galaxy simulations would miss.

For the $g^2T$ mock covariance, we repeat the shot-noise subtraction in every temperature realization, retaining the correlated uncertainties of the spectrum and the subtracted term (\cref{sec:shotnoise}).
\begin{equation}
 {\sf C}_{\rm mock}^{\nu\nu'}(L,L')=
 \operatorname{Cov}_{\rm sim}\!\left(
 \hat C_L^{g^2T_\nu}-\hat C_L^{g^2T_\nu,\,\mathrm{shot}},\,
 \hat C_{L'}^{g^2T_{\nu'}}-\hat C_{L'}^{g^2T_{\nu'},\,\mathrm{shot}}
 \right),
 \label{eq:paired_cov}
\end{equation}
where $\operatorname{Cov}_{\rm sim}$ is the sample covariance over the realizations. For $gT$, we use the sample covariance of the mock cross-spectra directly.

We check the errors of the null covariance $\widehat{\sf C}_0$ (\cref{sec:detect}), which uses the raw spectra before shot-noise subtraction, against the Gaussian (Knox) estimate \cite{Knox:1995dq}
\begin{equation}
 \sigma^2\big(C_L^{XT_\nu}\big)=
 \frac{\hat C_L^{XX}\hat C_L^{T_\nu T_\nu}
       +(\hat C_L^{XT_\nu})^2}
 {(2L+1)\,\Delta L\, f_{\rm sky}},\qquad X=g,\ g^2,
 \label{eq:knox}
\end{equation}
where the spectra are measured, $\Delta L=100$ is the bandpower width, and $f_{\rm sky}=\langle\mathcal M_g\mathcal M_\nu\rangle^2/\langle\mathcal M_g^2\mathcal M_\nu^2\rangle$, with $\langle\cdots\rangle$ here denoting a sky average.

\begin{figure*}[t]
\includegraphics[width=\textwidth]{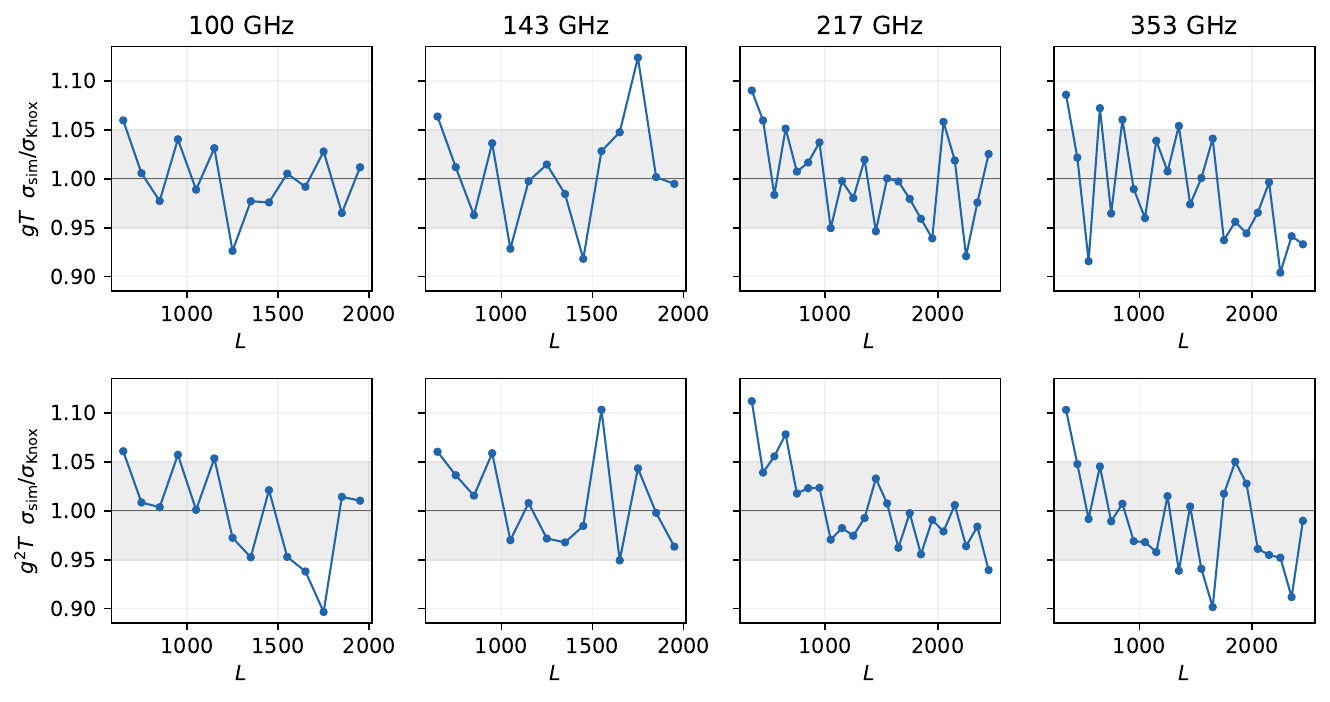}
\caption{Check of the null-covariance errors for $gT$ (top) and $g^2T$ (bottom) over the 72 analysis bandpowers. Each point is the error of the null covariance $\widehat{\sf C}_0$ (\cref{sec:detect}), computed from the raw spectra, divided by the Knox estimate of \cref{eq:knox}. The gray band marks ratios within 5\% of unity. The effect of ${\sf C}_{\rm halo}$ is described in \cref{sec:clusterterm}.}
\label{fig:errors}
\end{figure*}

\Cref{fig:errors} shows the ratio of the null errors estimated from the mocks to the Knox prediction over the 72 analysis bandpowers. The mean ratios are 0.998 for $gT$ and 0.999 for $g^2T$, showing agreement with unity. The 5\% scatter between bandpowers is close to the 4\% expected from estimating the errors with 300 realizations.

The simulations omit the cross-signal term $(\hat C_L^{XT_\nu})^2$, because their temperature realizations are independent of the galaxies, so ${\sf C}_{\rm mock}$ neglects it. Within the analysis ranges, this term contributes at most 0.35\% of the auto-spectrum product for $g^2T$ and 1.7\% for $gT$. Including it would lower the $gT$ S/N by at most 1\%.

\subsection{Halo-count covariance}
\label{sec:clusterterm}

The mocks keep the observed galaxy distribution fixed and therefore do not capture fluctuations in the halo population and its associated emission. An extra halo adds signal to several scales and frequency channels at once, correlating the bandpowers. We model this scatter by letting halo numbers and galaxy occupations fluctuate while holding the mean spatial profiles fixed. For a power spectrum, the resulting term approximates the one-halo mixed galaxy--temperature trispectrum contribution to the covariance \cite{COORAY_2002,Komatsu_2002}. For the filtered $g^2T$, we extend the same construction to the one-halo part of the connected mixed six-point function of four galaxy fields and two temperature fields, after shot-noise subtraction.

Let $c_L^{g^2T_\nu}(M,z)$ denote the mean weight of a halo of mass
$M$ and redshift $z$ in the one-halo bandpower. We calculate this
weight from \cref{eq:bisp_1h,eq:compression_h}, including the galaxy
filter, beams, pixel window, and bandpower window, so that
\begin{equation}
 C_L^{g^2T_\nu,\,1h}=
 \int\dd z\,\frac{\dd V}{\dd z\,\dd\Omega}
 \int\dd M\,\frac{\dd n}{\dd M}\,
 c_L^{g^2T_\nu}(M,z).
\end{equation}
Here $\dd n/\dd M$ is the comoving halo mass function, and $\dd V/(\dd z\,\dd\Omega)$ is the comoving volume per unit redshift and solid angle. For a uniform survey of solid angle $\Omega$, a single halo contributes $c_L^{g^2T_\nu}/\Omega$ to the measured bandpower. The galaxy part of this weight counts central--satellite and satellite--satellite pairs through $\langle N_cN_s\rangle$ and $\langle N_s(N_s-1)\rangle$, respectively (\cref{sec:profiles}). On large scales, this weight is proportional to $\langle N(N-1)\rangle$, where $N=N_c+N_s$ is the number of selected galaxies in the halo and $N(N-1)$ counts ordered distinct pairs.

To include occupation scatter, we rescale the mean weight of each halo by $N(N-1)/\langle N(N-1)\rangle$, with the expectation taken at fixed halo mass and redshift. This rescaling changes the amplitude of the halo signal but preserves its mean spatial profile. We also treat the masks $\mathcal M_g$ and $\mathcal M_\nu$ of \cref{sec:maps} as constant across the filtered signal of each halo, so that a halo $h$ with mass $M_h$, redshift $z_h$, sky position $\hat{\bm n}_h$, and galaxy count $N_h$ carries the single mask weight $\mathcal M_g(\hat{\bm n}_h)\mathcal M_\nu(\hat{\bm n}_h)$. Normalizing the sum by the mask-weighted solid angle $\int\mathcal M_g\mathcal M_\nu\,\dd\Omega$ gives
\begin{equation}
 \hat C_L^{g^2T_\nu,\,1h}\simeq
 \frac{1}{\int\mathcal M_g\mathcal M_\nu\,\dd\Omega}
 \sum_h\mathcal M_g(\hat{\bm n}_h)\mathcal M_\nu(\hat{\bm n}_h)\,
 \frac{N_h(N_h-1)}{\langle N(N-1)\rangle}\,
 c_L^{g^2T_\nu}(M_h,z_h).
 \label{eq:halo_sum}
\end{equation}
This approximation neglects the full mask-induced coupling between multipoles.

We treat halo numbers in small intervals of mass, redshift, and sky area as independent Poisson counts, with means set by the halo mass function and volume element, and we take galaxy occupations to be independent between halos. The covariance then averages the product of two contributions from the same halo. Each contribution scales with the pair count for $g^2T$ and with the galaxy count for $gT$, so averaging their product and dividing by the squared mean counts gives the occupation factors
\begin{equation}
 \begin{aligned}
 R_{g^2}(M)&=\left\langle\left[\frac{N(N-1)}{\langle N(N-1)\rangle}\right]^2\right\rangle
 =\frac{\langle[N(N-1)]^2\rangle}{\langle N(N-1)\rangle^2},\\[4pt]
 R_g(M)&=\left\langle\left[\frac{N}{\langle N\rangle}\right]^2\right\rangle
 =\frac{\langle N^2\rangle}{\langle N\rangle^2}.
 \end{aligned}
 \label{eq:occupation_cov}
\end{equation}
These factors are second moments rather than variances because halo numbers fluctuate together with the occupations. The occupation variance alone would omit the contribution from fluctuations in halo numbers. We evaluate the moments with the fixed HOD of \cref{sec:profiles}, which has Bernoulli centrals and conditionally Poisson satellites.

Integrating the products of the mask weights over sky position then gives the $g^2T$ halo-count covariance,
\begin{equation}
 \begin{aligned}
 {\sf C}_{\rm halo}^{\nu\nu'}(L,L';\bm\theta)&=
 \frac{1}{\Omega_{\rm eff}(\nu,\nu')}
 \int\dd z\,\frac{\dd V}{\dd z\,\dd\Omega}
 \int\dd M\,\frac{\dd n}{\dd M}\\
 &\quad\times R_{g^2}(M)\,
 c_L^{g^2T_\nu}(M,z;\bm\theta)\,
 c_{L'}^{g^2T_{\nu'}}(M,z;\bm\theta),
 \end{aligned}
 \label{eq:cluster_cov}
\end{equation}
where the effective solid angle
\begin{equation}
 \Omega_{\rm eff}(\nu,\nu')=
 \frac{\int\mathcal M_g\mathcal M_\nu\,\dd\Omega\;\int\mathcal M_g\mathcal M_{\nu'}\,\dd\Omega}
      {\int\mathcal M_g^2\mathcal M_\nu\mathcal M_{\nu'}\,\dd\Omega}
 \label{eq:omega_eff}
\end{equation}
accounts for the sky coverage and the overlap between channels, including the apodized mask weights. Replacing $c_L^{g^2T_\nu}$ and $R_{g^2}$ with $c_L^{gT_\nu}$ and $R_g$ gives the $gT$ covariance.

The weight $c_L$ sums the tSZ signal of each halo, including relativistic corrections, and its CIB signal, including self-dust. Because a fluctuation in halo number changes both signals together, the product of halo weights in ${\sf C}_{\rm halo}$ contains the tSZ and CIB variances and their covariance. Through these signals, ${\sf C}_{\rm halo}(\bm\theta)$ depends on the fitted emission parameters $P_0$, $c_{500}$, and $\eta_{\rm max}$, on $A_{353}^{g^2T}$ for $g^2T$, and on $\beta_{\rm tSZ}$ when it is free.

The predicted bandpowers use the sampled $M_{\rm eff}$, but the CIB contributions to ${\sf C}_{\rm halo}$ use fixed reference values, $\log_{10}(M_{\rm eff}/M_\odot)=12.67$ for $g^2T$ and 13.03 for $gT$. Recomputing the $g^2T$ covariance at the best-fit value, 12.985, and at 13.37 in the upper posterior tail changes the local Fisher estimates of the pressure uncertainties by at most 0.7\% and $-2\ln\mathcal L$ by less than 0.1.

In the baseline fits, ${\sf C}_{\rm halo}$ increases the $1\sigma$ uncertainty in $\log_{10}Y_{5R_{500}}$ by 53\% for $g^2T$ and 3\% for $gT$. It raises individual $g^2T$ bandpower errors by at most 8\%, but raises the uncertainty on the overall $g^2T$ amplitude by about 40--50\% at 100, 143, and 353~GHz. At the best fit, tSZ supplies almost all of ${\sf C}_{\rm halo}$, and the CIB, self-dust, and cross terms change the 353~GHz bandpower errors by less than 0.1\%. At 217~GHz, where the bandpass-integrated tSZ signal nearly vanishes even after relativistic corrections, the amplitude uncertainty grows by only about 0.1\%.

We do not evaluate the remaining non-Gaussian terms in full. For $gT$, we omit mixed galaxy--temperature trispectrum contributions beyond the one-halo approximation above. For $g^2T$, the full covariance also contains products of power spectra with mixed trispectra, products of bispectra, and connected six-point functions, of which we retain only the parts captured by ${\sf C}_{\rm mock}$ and the halo-count model. In particular, we omit their additional multi-halo contributions and the super-sample covariance from density fluctuations on scales larger than the survey.

\subsection{Likelihood and sampling}
\label{sec:inference}

We fit $gT$ and $g^2T$ separately, each to its 72 bandpowers across the four frequency channels, with the Gaussian likelihood
\begin{equation}
 -2\ln\mathcal L(\bm\theta)=
 \bm r^{\mathsf T}{\sf C}^{-1}(\bm\theta)\,\bm r+
 \ln\det{\sf C}(\bm\theta)+{\rm constant},\qquad
 \bm r=\bm d-\bm m(\bm\theta),
 \label{eq:complete_gaussian}
\end{equation}
where $\bm d$ contains the fitted bandpowers, $\bm m$ their model prediction, and ${\sf C}$ is given by \cref{eq:cov_split}. The determinant term is kept because ${\sf C}_{\rm halo}$ varies with the emission parameters.

We apply the Hartlap factor only to ${\sf C}_{\rm mock}$, because ${\sf C}_{\rm halo}$ is analytic and has no simulation noise. For Gaussian simulations and a Jeffreys prior on the true mock covariance, averaging the likelihood over the uncertainty in that covariance gives a distribution whose covariance is exactly \cref{eq:cov_split} \cite{Sellentin_2016,Heavens_2026}. Our Gaussian likelihood therefore has the correct covariance and misses only the heavier tails of the averaged likelihood.

Sampling, the best fit, and the goodness of fit all use this covariance. The goodness of fit at the best fit,
\begin{equation}
 \chi^2=\bm r^{\mathsf T}\left[h_{\rm H}^{-1}{\sf C}_{\rm mock}+{\sf C}_{\rm halo}(\bm\theta)\right]^{-1}\bm r,
 \label{eq:chi2_hartlap}
\end{equation}
is therefore the quadratic form of the sampled likelihood and needs no further correction. Rescaling ${\sf C}_{\rm halo}$ by $h_{\rm H}^{-1}$ as well changes $\chi^2$ by less than 0.1, because ${\sf C}_{\rm halo}$ mainly affects the overall signal amplitude, which the fit adjusts. The PTEs use $72-n_\theta$ degrees of freedom, where $n_\theta$ is the number of fitted parameters.

We sample the posterior with the flat priors of \cref{tab:g2params} using static nested sampling in \textsc{dynesty} \cite{Speagle_2020}. For each fit, we run the sampler twice with different random seeds, stopping each run when the estimated log-evidence remaining in the unexplored prior volume falls below $\Delta\ln Z=0.05$. Agreement between the two runs tests convergence, and we combine them into one posterior with twice the number of samples. The baseline fits use 1200 live points and draw new points by random-direction slice sampling. The free-slope $g^2T$ posterior follows a long, curved degeneracy (\cref{fig:betafree}), so for this fit we use 4800 live points and draw new points uniformly inside ellipsoids that enclose them. The medians and 16th and 84th percentiles of these two runs agree to within 2\% of the $1\sigma$ uncertainty for $\log_{10}Y_{5R_{500}}$ and 7\% for $\beta_{\rm tSZ}$. We report the maximum-likelihood sample as the best fit and summarize the posterior by its median and 16th--84th percentiles.

\section{Parameter constraints and comparison with previous work}
\label{sec:constraints}

\begin{figure*}[t]
\includegraphics[width=0.9\textwidth]{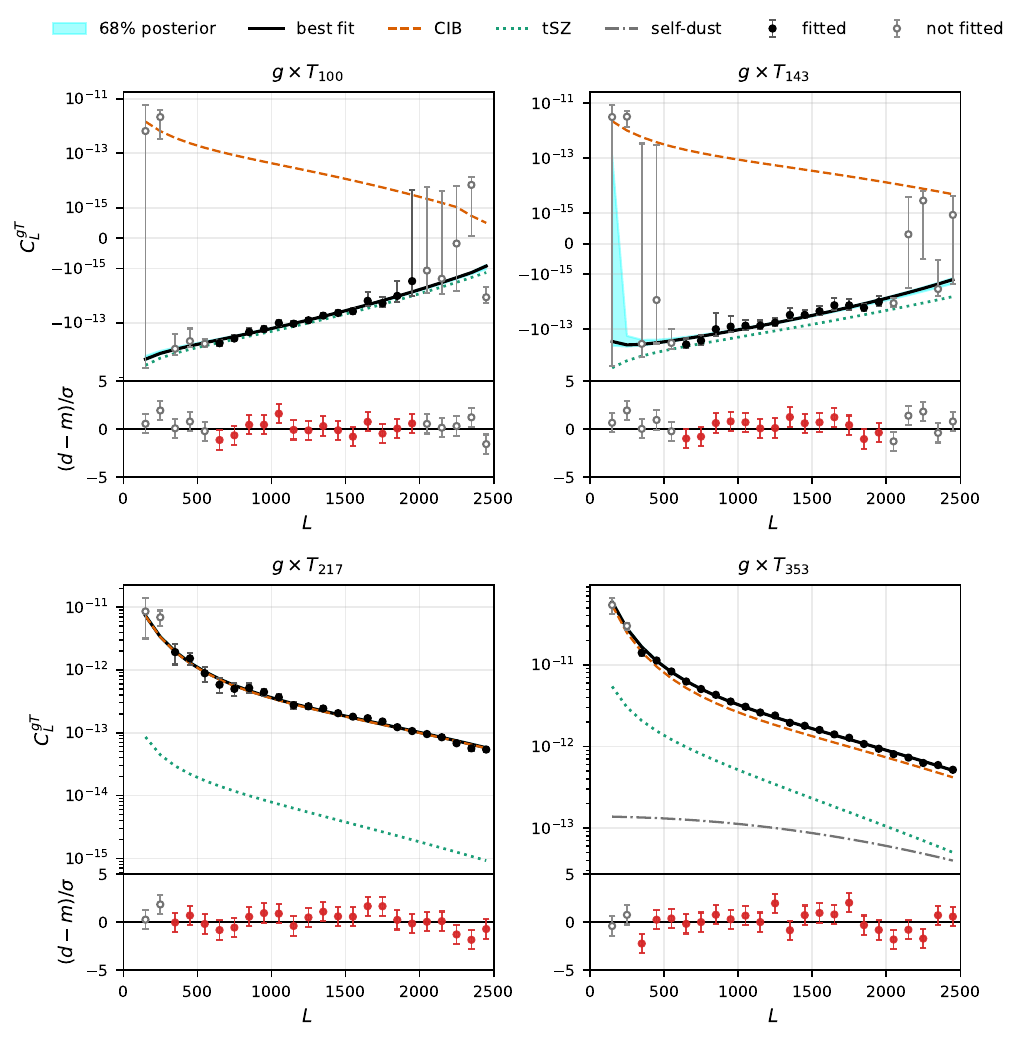}
\caption{As \cref{fig:g2t-fit}, for the baseline $gT$ fit.}
\label{fig:cross}
\end{figure*}

\begin{figure*}[!htbp]
\includegraphics[width=0.98\textwidth]{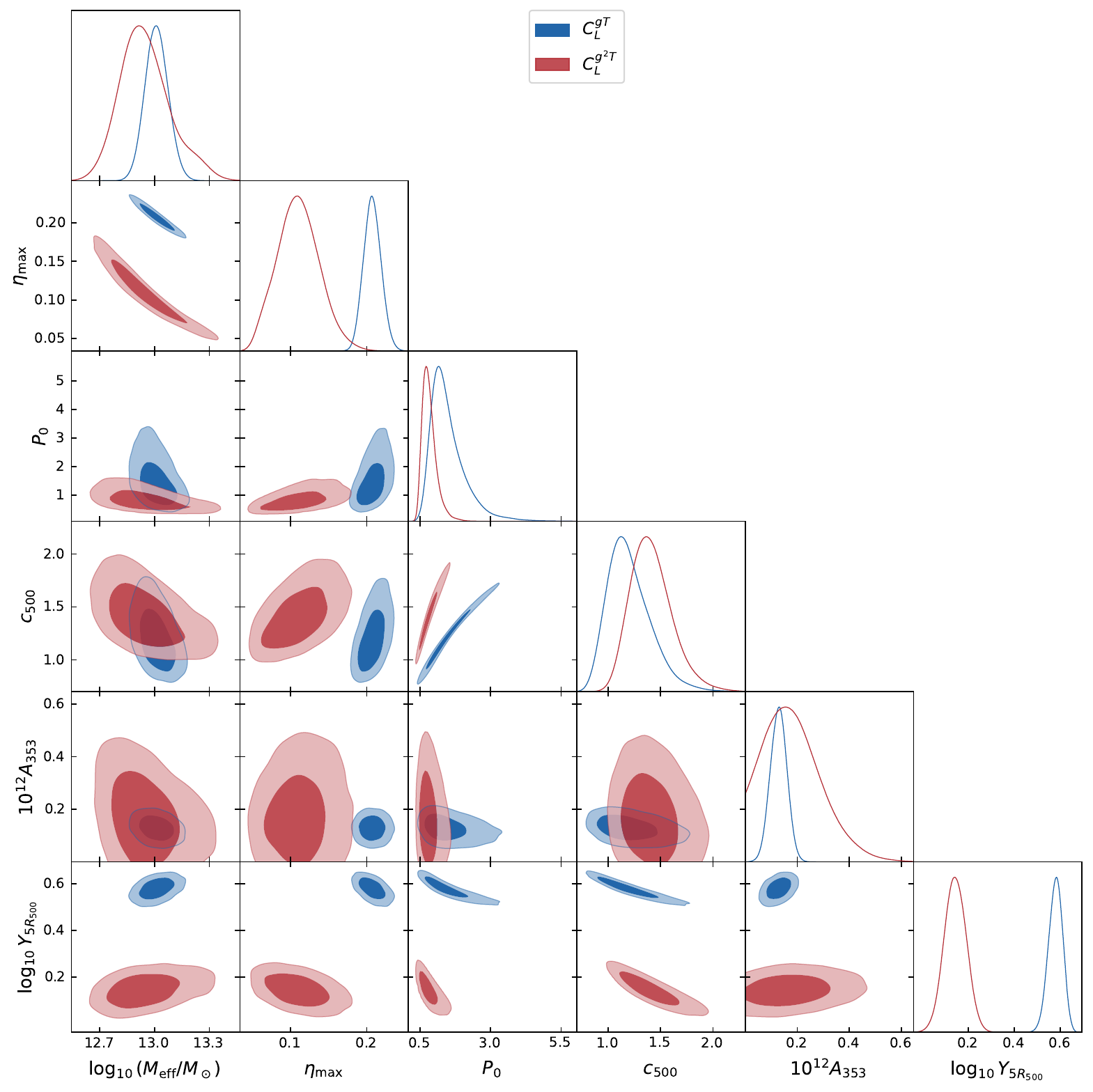}
\caption{Posteriors of the baseline $gT$ (blue) and $g^2T$ (red) fits for the shared free parameters and the derived $\log_{10}Y_{5R_{500}}$. The first two parameters describe the CIB and the next two the pressure profile. The fifth is the 353~GHz self-dust amplitude of each fit, $10^{12}A_{353}^{gT}$ or $10^{12}A_{353}^{g^2T}$ (\cref{eq:gt_cross_shot_model,eq:cib_coincident}).}
\label{fig:overlay}
\end{figure*}

This appendix compares the baseline fits of the two statistics and relates their constraints to previous analyses. \Cref{fig:cross} shows the $gT$ fit, and \cref{tab:g2params,fig:overlay} compare the posteriors of the two statistics. The error bars of $P_0$ and $c_{500}$ overlap between the fits, but the two posteriors occupy separate regions of the $P_0$--$c_{500}$ plane and therefore give different $Y_{5R_{500}}$. For the dust emission, the error bars of $M_{\rm eff}$ and of the self-dust amplitudes also overlap, whereas those of $\eta_{\rm max}$ do not. Among the dust parameters, only $\eta_{\rm max}$ has non-overlapping error bars.

\begin{table*}[t]
\centering
\renewcommand{\arraystretch}{1.3}
\begin{tabular}{l|c|cc|cc|c}
\hline
 & & \multicolumn{2}{c|}{$g^2T$ ($\chi^2_{\rm red}=1.15$)} & \multicolumn{2}{c|}{$gT$ ($\chi^2_{\rm red}=1.13$)} & \\
Parameter & Flat prior & Best fit & Posterior & Best fit & Posterior & Reference \\
\hline
$\log_{10}(M_{\rm eff}/M_\odot)$ & $[10,14]$ & $12.985$ & $12.939^{+0.145}_{-0.122}$ & $13.025$ & $13.011^{+0.064}_{-0.061}$ & $11.52^{+0.56}_{-0.71}$ \cite{Yan_2024} \\
$\eta_{\rm max}$ & $[0,1]$; $gT$: $[0.1,1]$ & $0.100$ & $0.109^{+0.029}_{-0.027}$ & $0.204$ & $0.207^{+0.012}_{-0.011}$ & $0.49^{+0.12}_{-0.16}$ \cite{Yan_2024} \\
$P_0$ & $[0,20]$ & $0.680$ & $0.785^{+0.279}_{-0.195}$ & $1.072$ & $1.333^{+0.703}_{-0.402}$ & $6.41$ \cite{2013Planck_tSZparams_bestfit} \\
$c_{500}$ & $[0.08,8]$ & $1.313$ & $1.395^{+0.212}_{-0.179}$ & $1.069$ & $1.174^{+0.231}_{-0.169}$ & $1.81$ \cite{2013Planck_tSZparams_bestfit} \\
$10^{12}A_{353}$ & $[0,2]$ & $0.182$ & $0.173^{+0.119}_{-0.100}$ & $0.138$ & $0.129^{+0.031}_{-0.032}$ & -- \\
$\log_{10}Y_{5R_{500}}$ (derived) & -- & $0.155$ & $0.144^{+0.048}_{-0.047}$ & $0.596$ & $0.580^{+0.029}_{-0.032}$ & $0.741^\dagger$ \cite{2013Planck_tSZparams_bestfit} \\
\hline
\end{tabular}
\caption{Flat priors, best fits, and posterior medians with $1\sigma$ uncertainties for the baseline $g^2T$ and $gT$ fits. Both fix $\beta_{\rm tSZ}=4.13$ and the other parameters specified in \cref{sec:profiles}. $A_{353}$ denotes each statistic's 353~GHz self-dust amplitude. Column headers give the best-fit reduced $\chi^2$ for 67 degrees of freedom. Reference values come from the M21 model of Ref.~\cite{Yan_2024}, with $M_{\rm eff}$ evaluated at $z=0$, and the \textit{Planck} pressure profile \cite{2013Planck_tSZparams_bestfit}. $^\dagger$Calculated from the published best-fit profile without propagating its uncertainty.}
\label{tab:g2params}
\end{table*}

For an external pressure reference, we integrate the \textit{Planck}--\textit{XMM-Newton} best-fitting profile of 62 nearby massive clusters \cite{2013Planck_tSZparams_bestfit} and obtain $\log_{10}Y_{5R_{500}}^{\rm ref}=0.741$.\footnote{We use the published best-fit parameters $(P_0,c_{500},\gamma,\alpha,\beta_{\rm tSZ})=(6.41,1.81,0.31,1.33,4.13)$ of Ref.~\cite{2013Planck_tSZparams_bestfit} and do not propagate uncertainty in the reference profile into this integral. That analysis finds that the dispersion of individual clusters about the mean profile dominates the statistical errors.} That fit combines X-ray profiles with stacked SZ measurements reaching about $3R_{500}$, so the integral to $5R_{500}$ partly extrapolates the fitted profile. At our adopted mass bias, $B=1.25$, the \textit{Planck}--\textit{XMM-Newton} profile gives 1.4 times the integrated pressure inferred from $gT$ and 4.0 times that from $g^2T$.

To compare with analyses that fit the mass bias at fixed pressure-profile shape, we keep the \textit{Planck}--\textit{XMM-Newton} profile shape fixed and calculate the mass bias $B_{\rm eq}$ that makes its physical integrated pressure match our $gT$ or $g^2T$ result. At fixed true halo mass and redshift, the physical pressure integral scales as (\cref{eq:p500})
\[
 P_{500}R_{500}^{3}Y_{5R_{500}}
 \propto Y_{5R_{500}}B^{-(5/3+0.12)}.
\]
Each model sets the integration aperture through its hydrostatic $R_{500}$. Matching the \textit{Planck}--\textit{XMM-Newton} profile to one of our fits therefore gives
\begin{equation}
 B_{\rm eq}=1.25\left(\frac{Y_{5R_{500}}^{\rm ref}}{Y_{5R_{500}}^{\rm fit}}\right)^{1/(5/3+0.12)}.
 \label{eq:beq}
\end{equation}
Here $Y_{5R_{500}}^{\rm ref}$ denotes the \textit{Planck}--\textit{XMM-Newton} integral, and $Y_{5R_{500}}^{\rm fit}$ is the value inferred from either statistic at our fixed $B=1.25$. Transforming our pressure posteriors gives median equivalent biases of $B_{\rm eq}=1.54\pm0.06$ for $gT$ and $2.70^{+0.17}_{-0.16}$ for $g^2T$. This conversion holds the galaxy occupation and the \textit{Planck}--\textit{XMM-Newton} profile fixed. It matches integrated pressure only, rather than the radial profile or angular spectra, and does not provide fitted mass-bias constraints.

\begin{figure*}[t]
\includegraphics[width=0.9\textwidth]{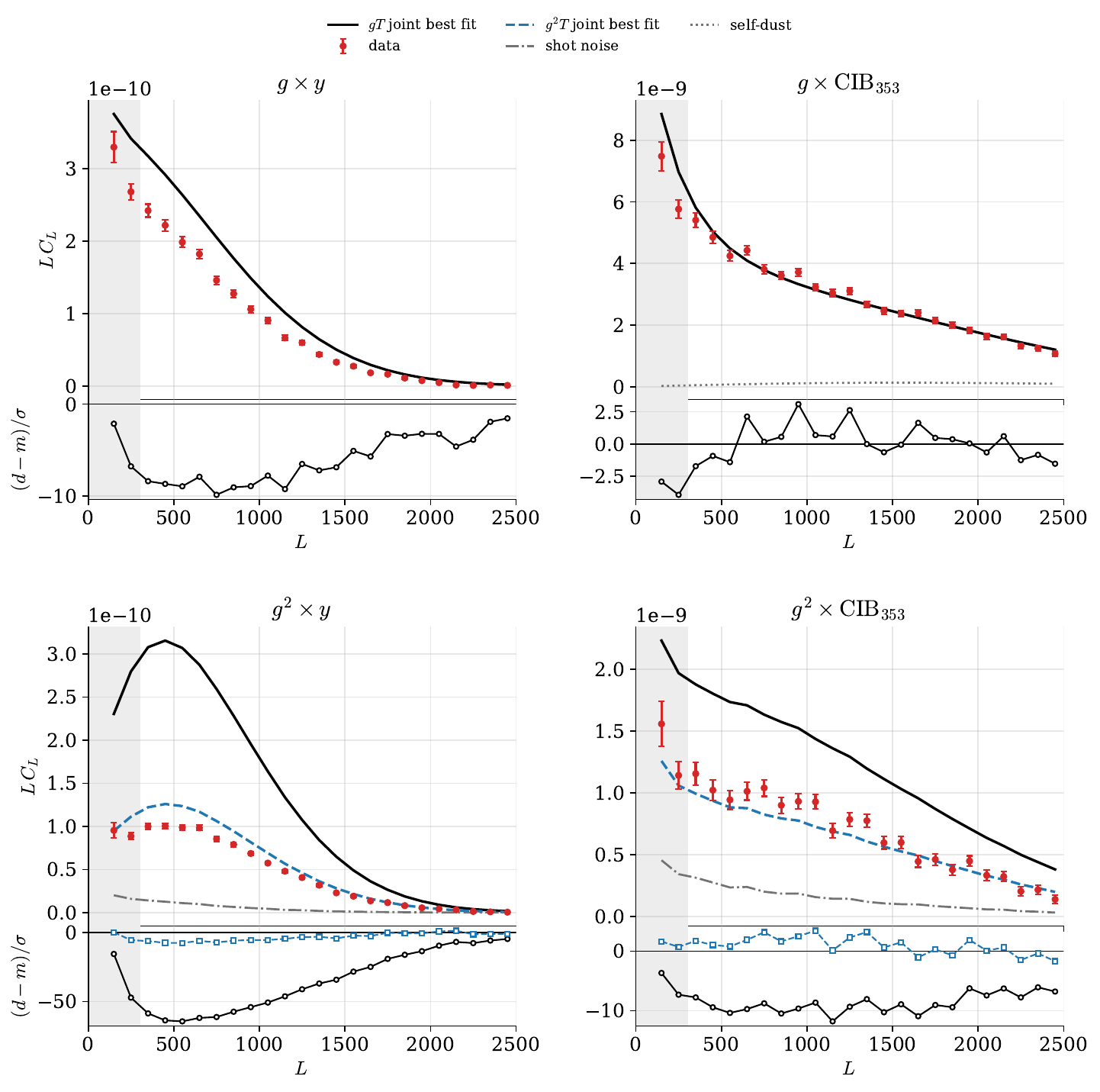}
\caption{Spectra with the NILC $y$ and 353~GHz CIB maps, compared with predictions from the baseline frequency-map fits. Black curves use the $gT$ best fit, and blue dashed curves use the $g^2T$ best fit in the bottom panels. CIB predictions include retained 353~GHz tSZ emission. Gray dot-dashed and dotted curves show shot noise and self-dust. Gray bands mark $L<300$, below the multipole ranges of the frequency-map fits. Lower panels show $(d-m)/\sigma$ using diagonal errors.}
\label{fig:component-validation}
\end{figure*}

Ref.~\cite{Ibitoye:2022oot} uses the same \textit{Planck}--\textit{XMM-Newton} profile and mass-bias definition. Its joint analysis of WISE galaxy clustering at $z\approx0.24$, the \textit{Planck} $y$ auto-spectrum, and their cross-spectrum, with a linear galaxy bias, gives $B=1.50\pm0.07\,({\rm stat})\pm0.34\,({\rm sys})$, close to our $gT$ equivalent value. Our NILC $g\times y$ amplitude is lower than the frequency-map prediction (\cref{fig:component-validation}), so this comparison also depends on whether we infer pressure from the frequency maps or the component-separated $y$ map.

Ref.~\cite{Koukoufilippas_2019} fits $y$-map cross-correlations with 2MPZ and WISE$\times$SuperCOSMOS galaxies at $z\lesssim0.4$, using an HOD and a free one-halo galaxy--pressure correlation. This freedom leaves the two-halo term to constrain the mass bias. Its corrected result \cite{Koukoufilippas_2020_erratum} is $1-b_{\rm H}=0.75\pm0.03$, equivalent to $B=1.33\pm0.05$. Combining the tSZ power spectrum with \textit{Planck} CMB constraints, Ref.~\cite{Bolliet_2018} finds $B=1.71\pm0.17$. These central estimates bracket our $gT$ equivalent value and lie below the $g^2T$ value. Differences in galaxy selection, redshift, and pressure modeling prevent a direct statistical comparison.

For the CIB parameters, two earlier analyses use the same star-formation-efficiency model. From the \textit{Planck} and \textit{Herschel} CIB power spectra, Ref.~\cite{Maniyar_2021} finds $\log_{10}(M_{\rm eff}/M_\odot)=12.94\pm0.02$ and $\eta_{\rm max}=0.42^{+0.03}_{-0.02}$. Its width parameters correspond to $\sigma_{M,0}^{\rm CIB}=1.75$, $z_c=1.5$, and $\tau\simeq1.00$ in our convention (\cref{eq:cib_width}). Our characteristic masses are close to this value, while both peak efficiencies are lower. This comparison depends on the efficiency-function width, which correlates with $\eta_{\rm max}$ and is fixed to $\sigma_{M,0}^{\rm CIB}=2.74$ in our fits.

From three unWISE samples cross-correlated with CIB maps at 353--857~GHz, the M21 model of Ref.~\cite{Yan_2024} gives $\log_{10}(M_{\rm eff}/M_\odot)=11.52^{+0.56}_{-0.71}$ at $z=0$, with no significant redshift evolution, and $\eta_{\rm max}=0.49^{+0.12}_{-0.16}$. Both of our fits favor higher characteristic masses and lower peak efficiencies, with uncertainties smaller by factors of 5--12. That analysis also varies the galaxy occupation, efficiency-function width, redshift evolution of $M_{\rm eff}$ and the infrared threshold, and additional nuisance parameters that our fits fix or omit. Correlations of $M_{\rm eff}$ with $\sigma_{M,0}^{\rm CIB}$, $\tau$, and its redshift evolution can account for much of the difference in precision. Our tighter uncertainties therefore remain conditional on the parameters we hold fixed. The analysis of Ref.~\cite{Yan_2024} draws its CIB constraints mainly from 545 and 857~GHz, whereas ours stops at 353~GHz, where we must separate CIB emission from tSZ.

\section{Robustness of the pressure difference}
\label{sec:robustness}

We test how frequency coverage, multipole cuts, and emission modeling affect the ratio of the $gT$ and $g^2T$ posterior medians of $Y_{5R_{500}}$. Across the tests in \cref{tab:robustness}, this ratio changes by $-6\%$ to $+10\%$ from its baseline value of 2.73.

\begin{table}[t]
\centering
\renewcommand{\arraystretch}{1.3}
\begin{tabular}{l|c|c|c}
\hline
Test & $g^2T$ & $gT$ & Ratio \\
\hline
Baseline & $0.144^{+0.048}_{-0.047}$ & $0.580^{+0.029}_{-0.032}$ & 2.73 \\
Without 353~GHz & $0.069^{+0.064}_{-0.069}$ & $0.516^{+0.054}_{-0.056}$ & 2.80 \\
$L_{\rm min}=400$ at 353~GHz & $0.124^{+0.046}_{-0.047}$ & $0.570^{+0.030}_{-0.032}$ & 2.79 \\
$L_{\rm min}=400$ at 100 and 143~GHz & $0.120^{+0.043}_{-0.044}$ & $0.598^{+0.023}_{-0.025}$ & 3.00 \\
\hline
Free $\beta_{\rm tSZ}$ & $0.164^{+0.060}_{-0.066}$ & $0.597^{+0.026}_{-0.029}$ & 2.71 \\
Self-dust in all channels$^\dagger$ & $0.120^{+0.046}_{-0.045}$ & $0.531^{+0.031}_{-0.030}$ & 2.58 \\
Radio emission & $0.152^{+0.046}_{-0.048}$ & $0.592^{+0.030}_{-0.031}$ & 2.75 \\
Magnification$^\ddagger$ & $0.139$ & $0.577$ & 2.74 \\
\hline
\end{tabular}
\caption{Posterior medians and $1\sigma$ uncertainties of $\log_{10}Y_{5R_{500}}$, and the ratio of the $gT$ and $g^2T$ medians of $Y_{5R_{500}}$. Baseline scale cuts follow \cref{sec:ranges}. Scale-cut rows modify only the named channels. All fits fix $\beta_{\rm tSZ}=4.13$, except the free-slope row. The radio row uses the Rayleigh--Jeans index $-2.8$, which gives the largest change in the ratio. $^\dagger$Self-dust follows the adopted CIB spectrum across all four channels. $^\ddagger$Shifts estimated at fixed profile shape and added to the baseline medians without refitting.}
\label{tab:robustness}
\end{table}

Removing the dust-sensitive 353~GHz channel lowers $Y_{5R_{500}}$ by 16\% for $g^2T$ and 14\% for $gT$ and broadens both posteriors, while their ratio changes by only 3\%. Removing the mask-sensitive 353~GHz bandpower at $300\leq L<400$ (\cref{sec:tests}), or adding the 100 and 143~GHz bandpowers at $400\leq L<600$, produces smaller shifts. The pressure ratio remains 2.73--3.00.

Allowing the outer pressure slope to vary also leaves the pressure difference nearly unchanged. With $\beta_{\rm tSZ}$ free (flat prior $[0.05,15]$), we find $\log_{10}Y_{5R_{500}}=0.164^{+0.060}_{-0.066}$ for $g^2T$ and $0.597^{+0.026}_{-0.029}$ for $gT$, a ratio of 2.71. The two statistics constrain the slope to $\beta_{\rm tSZ}=3.35^{+2.99}_{-0.62}$ and $2.85^{+0.76}_{-0.22}$, respectively, with $gT$ favoring a shallower outer slope than the \textit{Planck} reference. Ref.~\cite{Zhao_2026} also finds shallower slopes, $3.19^{+0.46}_{-0.36}$ and $2.54^{+0.15}_{-0.14}$, for unWISE samples at $\bar z=0.14$ and 0.23. Those fits vary only the pressure normalization and outer slope and use different fixed profile parameters and multipole ranges.

\begin{figure*}[!htbp]
\includegraphics[width=0.8\textwidth]{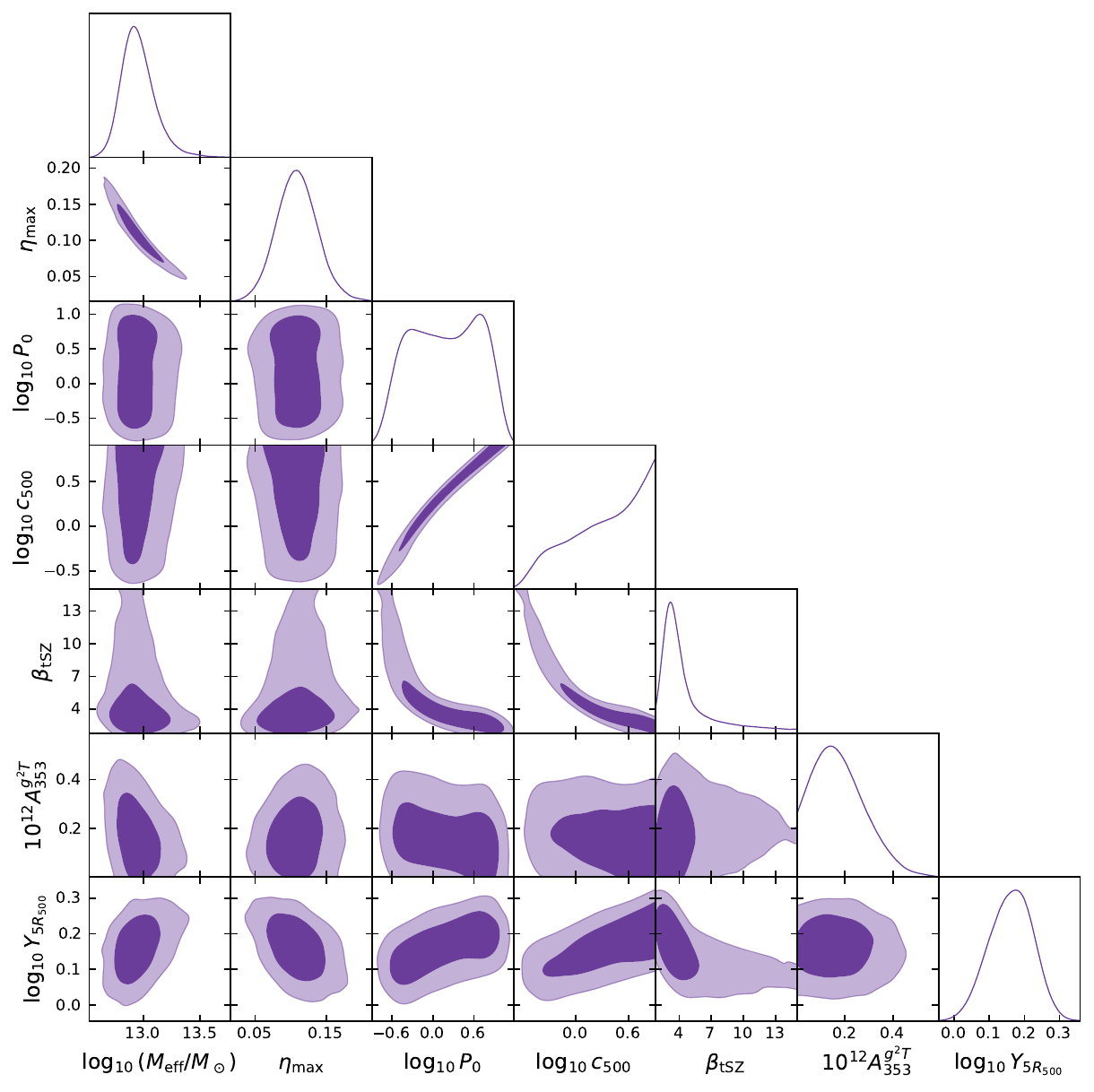}
\caption{As \cref{fig:g2t-corner}, with $\beta_{\rm tSZ}$ free. $P_0$ and $c_{500}$ have linear priors and logarithmic axes.}
\label{fig:betafree}
\end{figure*}

The free-slope $g^2T$ fit shows a strong degeneracy among $P_0$, $c_{500}$, and $\beta_{\rm tSZ}$ (\cref{fig:betafree}). Extending the upper limits on $P_0$ and $c_{500}$ from 20 and 8 to 50 and 16 raises the median $Y_{5R_{500}}$ by 6\%. This shift is small compared with the pressure difference, but the extended posterior still reaches the upper $c_{500}$ limit, so some prior dependence remains.

The baseline fits include self-dust only at 353~GHz. We first test this choice by adding an independent 217~GHz amplitude to $gT$, using the same prior as at 353~GHz. The inferred pressure barely changes, and the 217~GHz amplitude has a 95\% upper limit of 7.4\% of the fitted 353~GHz amplitude, below the 13\% predicted by the adopted CIB spectrum. We then refit each statistic with one free self-dust amplitude at 353~GHz and scale it to the other channels using the adopted CIB spectrum. The 217~GHz data favor a lower normalization in $gT$, shifting the posterior median of $10^{12}A_{353}^{gT}$ from 0.129 to 0.021. For $gT$, the reduced $\chi^2$ increases from 1.13 to 1.40 (${\rm PTE}=0.017$), mainly through the 217~GHz bandpowers. The scaled self-dust contributions at 100 and 143~GHz remain below the measurement uncertainties. For $g^2T$, the reduced $\chi^2$ changes only from 1.15 to 1.18. Both inferred pressures decrease, giving a ratio of 2.58 between their posterior medians.

Correlated radio emission could partly cancel the negative tSZ signal at 100 and 143~GHz \cite{Tucci_2011,De_Zotti_2009,Zhao_2026}. We fit a non-negative term with the scale dependence of the clustered 353~GHz CIB signal and Rayleigh--Jeans power-law indices $-2.1$, $-2.4$, and $-2.8$, corresponding to flux-density indices $-0.1$, $-0.4$, and $-0.8$, without a separate radio occupation or shot-noise term. All three indices give a best-fit radio amplitude of zero. Marginalizing over the positive amplitude allows a stronger tSZ signal and raises $Y_{5R_{500}}$ by about 2\% for $g^2T$ and 3\% for $gT$, changing their ratio by less than 1\%. Within the radio models tested here, contamination has little effect on the inferred pressures and does not explain their difference.

Ref.~\cite{Zhao_2026} detects radio emission at $9.5\sigma$ and $11\sigma$ in its lower-redshift samples, with indices $-2.18\pm0.24$ and $-2.62\pm0.26$. Those fits include the 30--70~GHz \textit{Planck} channels and ACT measurements to $\ell\simeq6000$, whereas ours use 100--353~GHz to $L=2500$ and a higher-redshift sample.

For the count slope $s=0.455$, the leading $g^2T$ correction comes from $2g\,g_\mu$, where $g_\mu$ denotes the magnification contribution to the galaxy overdensity. At the best fits, the corrections are 0.7\% for $gT$ and 1.2--1.4\% for $g^2T$, measured as rms ratios in each channel, with little scale dependence for $g^2T$. Their near proportionality to the galaxy terms allows an estimate at fixed profile shape without refitting. We estimate that $Y_{5R_{500}}$ decreases by 1.3\% for $g^2T$ and 0.7\% for $gT$, changing their ratio by less than 1\%.

\section{Comparison with tSZ and CIB maps}
\label{sec:components}

The \textit{Planck} NILC Compton-$y$ map has a 10-arcmin beam and its own Galactic and point-source masks, and it retains some CIB emission \cite{2016Planck_ymap}, which we do not model. Ref.~\cite{Lenz_2019} built the CIB map from \textit{Planck} PR3 data in the region with $N_{\rm HI}<2.0\times10^{20}\,{\rm cm^{-2}}$, by subtracting the SMICA CMB map \cite{2020_Planck_SMICA} and the Galactic dust traced by H\,\textsc{i} emission. Because the map retains the 353~GHz tSZ signal, we add the 353~GHz tSZ term of the model to each CIB-map prediction, neglecting the tSZ residual of the subtracted SMICA map. This term makes up 12\% of the predicted $g\times{\rm CIB}$, and 32\% and 23\% of the predicted $g^2\times{\rm CIB}$ for the $gT$ and $g^2T$ parameters. We apply the masks supplied with the CIB map, convert it to $\Delta T/\Tcmb$ with the 353~GHz unit conversion, and apply the 353~GHz color correction to the model \cite{2014_Planck_SpectralResponse,2014_CIB_colorcorr}.

To check our processing of the CIB map, we compare our $g\times{\rm CIB}$ bandpowers with the unWISE Blue measurement of Ref.~\cite{Yan_2024}, which uses the same map. After we remove the 5-arcmin beam of the CIB map and the pixel window, the two agree to within about 15\% over $300\lesssim L\lesssim2000$, with a median ratio of 1.02. Ref.~\cite{Kusiak_2023} measured the same correlation over a larger area of the map, with $N_{\rm HI}<2.5\times10^{20}\,{\rm cm^{-2}}$, and our bandpowers lie within 8\% of theirs over the same range, with a median ratio of 0.97.

We fix the model parameters at the $gT$ best fit and predict the cross-spectra $g\times y$ and $g\times{\rm CIB}$ and the skewed bispectra $g^2\times y$ and $g^2\times{\rm CIB}$ (\cref{fig:component-validation}). We also predict the two skewed bispectra from the $g^2T$ best fit. For each statistic, we quote the amplitude ratio, the factor by which the prediction must be scaled to best match the measured bandpowers, so that a ratio of one means agreement. For $g^2X$, where $X$ denotes either map, we approximate the shot noise by the uniform-weight limit of \cref{eq:shot_uniform}, evaluated with the measured $gX$ bandpowers. The ratios use diagonal errors and omit ${\sf C}_{\rm halo}$.

With the $gT$ parameters, the amplitude ratio is 0.73 for $g\times y$ and 0.35 for $g^2\times y$, and 1.00 for $g\times{\rm CIB}$ and 0.57 for $g^2\times{\rm CIB}$. With the $g^2T$ parameters, the skewed-bispectrum ratios are closer to unity, 0.84 for $g^2\times y$ and 1.11 for $g^2\times{\rm CIB}$.

\end{document}